\documentclass{cmslatex}
\usepackage[paperwidth=7in, paperheight=10in, margin=.875in]{geometry}
\usepackage{amsmath}
\usepackage{amssymb}
\usepackage{chemarr}
\usepackage{graphicx}
\usepackage{color}
\definecolor{myblue}{rgb}{0.2, 0.3, 1.0}

\newcommand{\p}{\partial}
\newcommand{\fr}{\frac}
\renewcommand{\theequation}{\arabic{section}.\arabic{equation}}

\begin{document}
\title{A Discrete Stochastic Formulation for Reversible Bimolecular 
	Reactions via Diffusion Encounter\thanks{}}

%For each author, make a block with the following four macros:

\author{Mauricio J. Del Razo\thanks{Department of Applied Mathematics, University of Washington, Seattle, WA 98195-3925. (maojrs@gmail.com)}
{}\and{Hong Qian \thanks{University of Washington, Seattle, WA 98195-3925. (hqian@u.washington.edu) }}}

%Use \thanks statements for acknowedgements of grants and
%support. They will appear below all the authors' addresses, so be
%specific about which author is thanking whom:

%\thanks{}

% Use the standard latex environments for theorems, etc. Here is one
% possible method of declaring them: It numbers all results by the
% section, and uses a common numbering system for the different
% environmentts.

%\date{Received date / Revised version date}

\pagestyle{myheadings} \markboth{Stochastic Formulation for Reversible Bimolecular 
	Reactions via Diffusion Encounter}{M. J. Del Razo \& H. Qian} \maketitle

% \large

\begin{abstract}
The classical models for irreversible diffusion-influenced 
reactions can be derived by introducing absorbing boundary conditions to over-damped 
continuous Brownian motion (BM) theory. As there is a clear corresponding stochastic process, 
the mathematical description takes both Kolmogorov forward equation for the 
evolution of the probability distribution function and the stochastic
sample trajectories. This dual description is a fundamental characteristic of stochastic processes and
allows simple particle based simulations to accurately match the expected statistical behavior.
However, in the traditional theory using the back-reaction boundary condition to model reversible reactions with 
geminate recombinations, several subtleties arise: It is unclear what the underlying 
stochastic process is, which causes complications in producing accurate simulations; and it 
is non-trivial how to perform an appropriate discretization for numerical computations. 
In this work, we derive a discrete stochastic model that recovers the classical models and their 
boundary conditions in the continuous limit. In the case of reversible reactions, we recover
the back-reaction boundary condition, unifying the back-reaction approach with those of current simulation packages. 
Furthermore, all the complications encountered in the continuous models become trivial in the discrete model. Our formulation 
brings to attention the question: With computations in mind, can we develop a discrete reaction kinetics model 
that is more fundamental than its continuous counterpart?
\end{abstract}



\section{Introduction}\label{intro}

The theory of Brownian motion \cite{einstein1956investigations,qian2014statistics,uhlenbeck1930theory}, 
as developed by Einstein, Smoluchowski, and Langevin, describes the random motion of a particle immersed 
in a fluid resulting from its collisions with the atoms or molecules of the fluid.  There are two distinct 
physical effects from the collisions: A mean frictional force that resists a macroscopic motion
and a random fluctuating force, with zero mean, that rapidly changes the directions of movements.  As 
with any model, the theory itself is
a mathematical idealization of the physical reality. For instance, as a continuous function
of time, overdamped Brownian motion has a fractal geometry, i.e. between any two instants in time, 
there is an infinite amount of fluctuations and changes of direction of the particle. However, any 
physical process that follows Newtonian mechanics always has two times sufficiently 
close that a particle only moved on a straight line, known as mean free path.

In the current work, we deal with reversible chemical reactions
mediated by diffusion in aqueous solution, which is a complex problem 
built over the Brownian Motion framework. Historically, 
Smoluchowski, and Collins and Kimball \cite{collins1949diffusion,smoluchowski1917versuch} formulated 
a macroscopic theory for the irreversible formation of a chemical species $C$ from separated $A$ and 
$B$ compounds in terms of diffusive motions of the reactants, leading to their encounter and chancy transformation.  
Interestingly, the stochastic process underlying the same diffusion equation and boundary
condition, known as {\em partially reflected Brownian motion} (PRBM) \cite{grebenkov2006}, 
predetermines an associated process of repeated futile encounter known as 
a gemination process \cite{agmon1990theory,andrews2004stochastic,berg1978diffusion,del2014fluorescence}, which
plays a central role in both association and dissociation processes. The rigorous mathematical problem 
of PRBM, however, is non-trivial \cite{atakis-nguyen-2013}.
It is best understood as the limit of a discrete random walk \cite{burdzy-chen-2008}, 
as explained in detail in the Appendix \ref{sec:appB}.

In practice, most complex problems in applied mathematics are 
inevitably solved computationally. The continuous mathematical descriptions 
of Brownian motion problems like the above are eventually 
discretized into algorithms that are appropriate for numerical computations. 
One therefore naturally seeks a formulation of applied ``Brownian motion'' 
problems directly in terms of a discrete representation. Such an approach, 
without loss of accuracy and rigor, bypasses two difficult mathematical subjects altogether: 
continuous stochastic path as the limit of a discrete Markov process
and numerical accuracy of high-order discrete algorithm for a continuous
problem. The objective of the present work is to follow this approach to develop a discrete stochastic 
model for reversible bimolecular reactions via diffusion encounter.  

It is important to mention that at this point we are not concerned in making a judgment 
of whether the physical reality is Newtonian or not but on a more computationally accessible approach.
Our premise is that all phenomena we observe as the physical reality at 
some time and space scales, in applied mathematics, can be formulated and implemented 
as discrete models, which is not necessarily the same as implementing discretization of continuous
physical laws. In 1960s computational fluid dynamics (CFD) had a similar problem emerged
that was first solved by Godunov in his 1959 revolutionary paper \cite{godunov1959difference}.
Before Godunov's work, the CFD numerical methods for compressible flow were plagued with difficulties.
His approach to solve these issues was to provide a discrete conservation law instead 
of a discretization of a conservation law solely \cite{godunov1959difference,randysrbook,raynal1972computing,torosbook}. 
The model presented herein shares that inspirational spirit; we will 
formulate a discrete stochastic process instead of a discretization of a continuous stochastic
process.

Additional motivation lies in the growing need for quantitative mathematical models, and fast algorithms, 
of small biological systems like cells or subcellular compartments of the cell. Most of these systems are
non-homogeneous in space and have a low number of molecules, so its modeling is based on
stochastic reaction-diffusion theory at mesoscopic scales. However, unlike the case of 
homogeneous reaction theory \cite{bartol1991monte,qian2011nonlinear}, the connection between the reaction-diffusion phenomena at 
different microscopic, mesoscopic and macroscopic scales is still a matter of recent research 
\cite{arnold1980consistency,feng1996hydrodynamic,hellander2014reaction,isaacson2008relationship,
isaacson2013convergent}. Our model contribution  
in this front is to unify the theory of reversible diffusion-influenced reactions 
\cite{agmon1984diffusion,kim1999exact,schurr-1970} with the different approaches to model 
reversible reactions taken by several simulation packages , like Smoldyn, FPKMC, eGFRD 
\cite{andrews2004stochastic,donev2010first,takahashi2010spatio} and others 
\cite{drawert2012urdme,hattne2005stochastic,schoneberg2013readdy, tomita1999cell, van2005simulating, wils2009steps}.
We also note that in a macroscopic phenomenological 
reaction-diffusion theory the diffusion coefficients of $A$ and $B$ and their association rate 
constant are three independent parameters \cite{fcs:prl,murrayjdbook}.  Mechanistically, however, 
the last is a function of the former two \cite{del2014fluorescence,smoluchowski1917versuch}. Our 
discrete model could also provide a guideline on how to establish these connections for nonlinear reversible reactions. 

With the discussion above in mind,  we begin with a general approach in the spirit of 
Langevin \cite{langevin1908theorie}. Consider the stochastic dynamic equations for an 
immersed particle of mass $m$,
\begin{align}
 dx= vdt,  \  \    m dv = \big( -\eta v + \xi(t) \big) dt;
 \label{:Langevin1}
\end{align}
where $x$ is the position, $v$ the velocity, $\eta$ the damping coefficient, $\xi(t)$ the 
white noise term that satisfies $\xi(t) dt =\sqrt{2k_B T\eta}\ dW_t$ with $W_t$ the 
standard Brownian motion or Wiener process, $k_B$ the Boltzmann constant and $T$ the 
temperature. The Wiener process satisfies $E[W_t] = 0$ and 
$E[W_t W_s] = \mathrm{min}(t,s)$, which implies $E[\xi(t) \xi(s)]=2k_BT\eta\delta(t-s)$. The stochastic 
trajectory in its integral form is given by
\begin{equation}
 x(t) = x(0) + \int_0^t e^{-\eta s/m} \left(v(0) + \fr{1}{m}\int_0^s e^{\eta \tau/m} \xi(\tau)d\tau\right)ds,
\end{equation}
and its probability distribution dynamics is described by the multivariate Fokker-Planck equation, 
which in this specific case, it is usually referred as the Klein-Kramers equation, 
\begin{equation}
 \fr{\p \mathit{f}(x,v,t)}{\p t} = \fr{\p}{\p v}\left[ 
 \left(\fr{k_B T \eta}{m^2}\right)\fr{\p f}{\p v} + \fr{\eta v}{m}f\right] - v \fr{\p f}{\p x}.
\end{equation}

In the overdamping limit of equation (\ref{:Langevin1}), we obtain $dx = \fr{1}{\eta}\xi(t) dt,$
which yields a simpler Fokker-Planck equation
\begin{equation}
 \fr{\p f(x,t)}{\p t} = D \fr{\p^2 f(x,t)}{\p x^2},
\label{heat-eq}
\end{equation}
with the diffusion coefficient given by the Einstein relation: $D=k_B T/\eta$. 
This equation describes the probability distribution dynamics of standard Brownian 
Motion or Wiener process with no drift. Note it can be extended to varying damping 
coefficient $\eta(x)$, which would make the diffusion coefficient not constant, and 
it will change the form of the equation.  In particular, the issue of It\={o} vs.
Stratonovich or divergence form of the diffusion term matters \cite{aoping-2007}.

Although Eq. \ref{heat-eq} has an identical form as the classical
diffusion equation for a density of particles, it has a much fundamental
character; it does not rely on Fick's law.  The classical equation should be 
understood as the equation of {\em mean density} of a large number of identical, 
independent Brownian particles.  Fick's law then is an emergent statistical phenomenon. 

The theory of chemical reaction in aqueous solution, mediated by diffusion encounter, is based 
on a three-dimensional version of the standard Brownian motion we just 
presented, with the addition of an absorbing or partially absorbing boundary condition 
to model the ``event of encounter'' in a reaction; it is the core of Smoluchowski's 
theory  \cite{collins1949diffusion,smoluchowski1917versuch}. However, as Brownian 
Motion is such a powerful and useful mathematical idealization, it is easy to assume that 
it is a first principle. For instance, we tend to think of random walks as approximations 
to standard Brownian motion (or more accurately to a Wiener Process), immediately 
assuming Brownian motion is the best description of the physical world. In reality, 
the theory of Brownian motion is a mathematical limit, and one could even argue Random 
Walks might provide models closer to what one observes in reality than Brownian motion 
theory. With computational tasks in mind, why shouldn't we think standard Brownian motion 
theory is the one providing an approximation to some specific kind of Random walks? 

For relatively simple processes, as in diffusion-influenced irreversible reactions, 
standard Brownian motion provides a robust theoretical framework. However, when 
dealing with more complex processes like reversible reactions, a great
deal of subtleties arise 
\cite{agmon1984diffusion,andrews2004stochastic,berg1978diffusion,kim1999exact,prustel2012exact}. 
Many of the issues one encounters while trying to model more and more complex 
processes are intrinsic to the fact that one had adopted a specific mathematical 
idealization, especially when dealing with such an abstract idealization like Brownian
Motion. For instance, it has been shown that the continuous diffusion approximation of  
discrete reaction networks can fail to represent correctly the mesoscopically interesting
steady-state behavior of bi-stable systems \cite{vellela2009stochastic}, an issue of utter relevance
in biochemical cell dynamics. The purpose of this paper is to convince the reader that an alternate discrete 
approach to reversible stochastic reaction-diffusion might provide a simpler,  more 
robust, and computation-friendly framework where these subtleties are no longer an 
issue. It also unifies previous theoretical approaches
with more recent simulation algorithms, contributing to a better understanding of 
reversible stochastic reaction-diffusion. Although the model here presented is for a 
relatively simple problem, it provides the guidelines for a different line of research
that could address fundamental issues in stochastic reaction-diffusion theory and simulations.
In the next subsection, we present a condensed review of some of the classical 
models for bimolecular reactions that are mediated by diffusion encounter.

\subsection{Bimolecular reactions mediated by diffusion encounter}\label{diffencounter}

{\bf\em Diffusion-influenced reactions.}
Consider the reaction $A+B\overset{k_S}{\longrightarrow} C$, where $A$ and 
$B$ are two reactive spheres diffusing in space. Fix the frame of reference 
at the center of $A$ and assume many $B$ particles diffuse around $A$ with a diffusion 
coefficient given by the sum of $A$ and $B$ diffusion coefficients, $D=D_A+D_B$, 
following standard Brownian motion. Whenever the $B$ molecules reach by diffusion 
the boundary $\sigma$ given by the sum of the radii of $A$ and $B$, $\sigma = R_A + R_B$, 
we assume a reaction occurs. We call this reaction a purely diffusion-controlled reaction. 
Smoluchowski classical work, given in detail in 
\cite{collins1949diffusion,shoup1982role,smoluchowski1917versuch}, 
calculates the association rate constant $k_S$.\footnote{In fact, how to
define the association rate constant $k_S$ is itself an important issue.  
Ideally, if the waiting time distribution for successive reactions is exponential, 
then a single rate parameter $k_S$ suffices.  When the distribution is non-exponential, 
usually one takes the reciprocal of the mean time as the $k_S$, which can be shown to 
agree with an appropriate steady-state flux.  The spatial dimension
has a crucial role in this problem, which is certainly related to  P\'{o}lya's recurrence theorem.}

In Smoluchowski's original work, the concentration of molecule $B$
surrounding the $A$ is denoted by $c(r,t)$. It obeys a simple 
three-dimensional diffusion equation,
\begin{subequations}
\begin{gather}
  \frac{\p c(r,t)}{\p t} = \nabla \cdot \big[D(r)  \nabla c(r,t)\big],
        \   \     r\in [\sigma,R] \label{eq:smol} \\[3mm]
  c(\sigma,t) = 0, \   \    c(R,t) = c_0.
\label{smolBC}
\end{gather}
\end{subequations}
As the problem is spherically symmetric, we obtain 
$\frac{\p c(r,t)}{\p t} = \frac{D}{r^2}\frac{\p}{\p r}\left(r^2 \frac{\p }{\p r} c(r,t)\right)$, for constant $D$.
The absorbing boundary condition at $\sigma$ represents the purely diffusion-controlled reaction 
with 100\% reaction for each and every encounter; the boundary condition at $R$ provides a 
bath of $B$ molecules that guarantees a constant concentration, $c_0$, at a distance $R \gg \sigma$.  
The quantity $c_0$ is identified as the ``bulk concentration'' in an aqueous solution containing many $A$'s 
and $B$'s. In the limit of $R \rightarrow \infty$, the time-dependent and
stationary solutions to this equation are given respectively by
\begin{align*}
  c(r,t) = c_0\left[1-\fr{\sigma}{r}\mathrm{erfc}\left(\fr{r-\sigma}{\sqrt{4Dt}}\right)\right], \ \ \ \
  c_{ss}(r) = c_0\left[1-\fr{\sigma}{r}\right].
\end{align*}
The steady state diffusion controlled association rate constant is then obtained from the flux at 
the reaction radius, $\sigma$, as
\begin{align*}
  k_S =\fr{4\pi D \sigma^2 c_{ss}'(\sigma)}{c_0} = 4\pi D \sigma.
\end{align*}
In the one and two dimensional case, it is worth mentioning that it is not possible to
obtain reaction rates in the same manner since the mean first passage times diverge as 
$R \rightarrow \infty$. There is a large literature on how to deal with this physically very 
different and mathematically challenging problem \cite{prustel2012exact,prustel2013theory,prustel2014rate,saffman1975brownian}.
Also, as pointed out in \cite{collins1949diffusion}, the time-dependent flux becomes 
infinite at $t=0$, which is unacceptable when the initial rate may be very significant. 
These issues, among others pointed out
in \cite{collins1949diffusion}, are weaknesses of the theory.

Collins and Kimball gave an improvement over Smoluchowski's 
theory \cite{collins1949diffusion}, in which the absorbing boundary 
at $\sigma$ is replaced with a partially absorbing boundary condition.
This is obtained by making the flux equal to the concentration at the reaction radius,
\begin{align}
  \left. 4 \pi \sigma^2 D \fr{\p c(r,t)}{\p r} \right|_{r=\sigma} = \kappa c(\sigma,t), \label{colBC}
\end{align}
where $\kappa$ controls the degree of diffusion-influence in the reaction 
($\kappa=0$ means no reaction, while $\kappa \rightarrow \infty$ means the 
reaction is diffusion limited).  In stochastic simulations, $\kappa$ is 
intimately related to the probability of being absorbed ($p$) or being
reflected. But the relation is non-trivial as signified by the dimension
of $\kappa$: [length]$^3$[time]$^{-1}$.  In fact, both $D$ and $\kappa$
are best understood in terms of a discrete setting: 
$D= \frac{\Delta x^2}{6\Delta t}$ and $\frac{\kappa}{4\pi\sigma^2}=
\frac{\Delta x}{\Delta t}p$, see Appendix \ref{sec:appB}. 

Solving Eq. \ref{eq:smol} with $\frac{\p c(r,t)}{\p t}=0$ and with this boundary 
condition yields the steady state, and the partially absorbing reaction rate of Collins and Kimball,
\begin{eqnarray}
 c_{ss}(r) &=& c_0\left[1-\fr{\kappa \sigma}{4\pi D \sigma + \kappa} 
 \left(\fr{1}{r}\right)\right], \label{eq:cksol} \\[2mm]
  k_{CK} &=& \fr{\kappa k_S}{\kappa + k_S}.
  \label{eq:ck01}
\end{eqnarray}
Note that if $\kappa \rightarrow \infty$ we recover the purely diffusion limited 
reaction rate $k_S$. Furthermore, in the full time dependent solution with the new boundary 
condition, the flux will no longer be singular at $t=0$ \cite{collins1949diffusion}. 

\begin{figure}
\centering
\includegraphics[width=0.4\columnwidth]{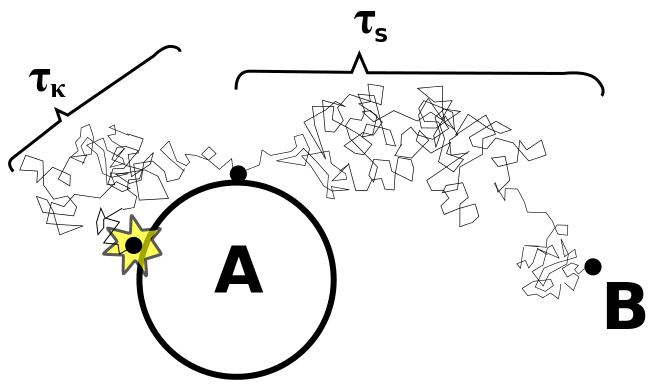}
\caption{Interpretation of the times corresponding to the inverse of the reaction 
rates in the Collins and Kimball model. Note that after a time $\tau_S$ when particle 
$B$ collides with $A$, the reaction might not happen; it still has to wait a time 
$\tau_{\kappa}$. This is the stochastic trajectory interpretation of a partially absorbing 
boundary. In a more formal interpretation PRBM should be understood as purely reflective 
Brownian Motion conditioned to stop at a random time \cite{burdzy-chen-2008,grebenkov2006}, 
see Appendix \ref{sec:appB}.}
\label{fig:FPTs}
\end{figure}

The reaction rates defined through steady-state flux, given in Eq. \ref{eq:ck01}, have a clear
probabilistic interpretation in terms of the mean passage times, $\tau$
\cite{shoup1982role,szabo1980first}.  We can assume 
$\tau_{CK} = \tau_S + \tau_{\kappa}$, where $\tau_{CK}$ is the mean passage time until the first 
reaction occurs in the Collins and Kimball model, $\tau_S$ is Smoluchowski's mean first 
passage time from a uniform distribution outside the absorbing boundary to the absorbing boundary, 
and $\tau_{\kappa}\sim\kappa^{-1}$ is 
the mean passage time starting from the absorbing boundary until the molecule is bound, 
this is illustrated in Figure \ref{fig:FPTs}. 
Note Eq. \ref{eq:ck01} is immediately recovered by using the inverse relation $\tau_{CK} \sim 1/k_{CK}$.   
A reaction is reaction-probability limited if $k_S\gg\kappa$, such that $k_{CK}\approx \kappa$.  
Therefore, Collins and Kimball's theory is applicable 
to cases ranging from diffusion-limited scenario $k_S\ll\kappa$ to  
reaction-probability-limited scenario.

The reaction diffusion problems framed as above can equivalently be described in terms of the Green's functions for an isolated pair of 
$A$ and $B$ molecules \cite{van2005green,van2005simulating}. 
Consider again one molecule $A$ fixed in the origin. We will denote $f(r,t|r_0)$ 
the probability of a particle $B$ being a distance $r$ from $A$ at time $t$ given 
that it was at $r_0$ at time $0$. This transition probability will obey the diffusion 
equation, initial condition and boundary condition,
\begin{subequations}
\label{eq9}
\begin{gather}
  \frac{\p f(r,t|r_0)}{\p t} = \nabla \cdot\big[ D(r)  \nabla f(r,t|r_0) \big].\label{eq:smol2} \\[3mm]
  f(r,0|r_0) = \fr{\delta(r-r_0)}{4\pi r_0^2}. \\[3mm]
  \lim_{r\rightarrow\infty}f(r,t|r_0) = 0,
\end{gather}
\end{subequations}
and an extra boundary condition at $r=\sigma$, which can be Eq. (\ref{smolBC}) or  Eq. (\ref{colBC}) 
written in terms of $f(r,t|r_0)$. Note this is an equation for the probability $f(r,t|r_0)$, which is 
the ``remaining'' probability density function in the presence of an absorbing or partially absorbing 
boundary (diffusion with killing). Then
\[
-\frac{d}{dt}\int_{\sigma}^{\infty} 4\pi r^2 f(r,t|r_0) dr
\]
is the probability density function for the absorbing time. \\

{\bf\em Reversible bimolecular reaction via diffusion encounter.}
The Green's function formulation for an isolated pair has been extended to modeling reversible reactions 
in one, two and three dimensions \cite{agmon1984diffusion, kim1999exact, prustel2012exact}.  For spherical symmetry and constant $D(r)=D$, 
these extensions consist of augmenting boundary 
condition (\ref{colBC}) with a ``back-reaction'' with rate $\mu$:
\begin{align}
 \left. 4 \pi \sigma^2 D \fr{\p f(r,t|r_0)}{\p r} \right|_{r=\sigma} = 
 \kappa f(\sigma,t|r_0) - \mu\big[1 - S(t|r_0)\big], 
\label{backBC} 
\\[3mm]
 \mathrm{where} \hspace{2mm} S(t|r_0) = 1 - \int_0^t 4\pi \sigma^2 D 
 \left. \fr{\p f(r,\tau|r_0)}{\p r}\right|_{r=\sigma} d\tau, \nonumber
\end{align}
and $S(t|r_0)$ is referred as the survival probability in the literature. 
An analogous version of the boundary condition in (\ref{backBC}) first 
appeared in \cite{schurr-1970}, which also made an important statement that in a chemical
reaction equilibrium, the $\lim_{t\rightarrow\infty} f(r,t|r_0)$ should be
uniform in space independent of $r$.

The notion of survival probability doesn't really make sense when talking about reversible reactions. A more 
appropriate name for $S(t|r_0)$ is the probability of $B$ not being bound at time $t$ given it initially 
was a distance $r_0$ from $A$, regardless if the particle was bounded at some time between $0$ and $t$. 
This is clear from the fact that 
\begin{align}
S(t|r_0) =& 
 1 - \int_0^t  4\pi \sigma^2 D \left. \fr{\p f(r,\tau|r_0)}{\p r}\right|_{r=\sigma} d\tau \label{eq:survprob}  \\
 =& \int_\sigma^{\infty} 4\pi r^2f(r,t|r_0) dr, \nonumber
\end{align}
which is easily proved by differentiating Eq. \ref{eq:survprob} by $t$, then using Eq. \ref{eq:smol2} and 
the boundary conditions, then repeat the inverse process and evaluate at $t=0$ to 
find the integration constant to be one. The second integral is clearly
the probability of being unbound as we mentioned before. The exact solution of this 
problem is given in \cite{kim1999exact}. In \cite{popov2001three,popov2001three2}, 
the authors implemented numerical simulations using this model as a starting point. 

If one restricts the diffusion of $B$ in a finite space and replaces Eq. \ref{eq9}c 
with a reflecting boundary at $R$, then the stationary equilibrium solution to 
Eq. \ref{eq9}a with boundary condition Eq. \ref{backBC} gives
$f_{eq}(r)\equiv f_{eq}(\sigma) = (\mu/\kappa) p_C$ in which
$p_C$ is the equilibrium probability of $C$: $p_C = 1- f_{eq}(\sigma)V_B$,
where $V_B=\frac{4\pi}{3}\big(R^3-\sigma^3\big)$ is the volume
available to the diffusing $B$ \cite{schurr-1970}.
Therefore, the theory provides a rigorous {\em equilibrium
constant} for the reversible association reaction
\begin{equation}
           K_{eq} \equiv \left(\frac{p_C}{1-p_C}\right)V_B
                        = \frac{\kappa}{\mu}.
\label{eq12}
\end{equation} 
This implies that if the bimolecular association rate constant is
$\kappa$, then the unimolecular dissociation rate constant has to be $\mu$.
On the other hand, if we choose Eq. \ref{eq:ck01} as the on-rate constant 
$k_{\mathrm{on}}=\kappa k_S/(k_S+\kappa)$, as suggested by Fig. \ref{fig:FPTs}, then the off-rate 
constant has to be $k_{\mathrm{off}}=\mu k_S/(k_S+\kappa)$. The validity of Eq. \ref{eq12} assumes two 
well-defined states, $A+B$ and $AB\equiv C$, with Markov transition in between. Strictly 
speaking, this requires the dynamics within each of the two states to be sufficiently rapid while the
transitions between the two states as relatively rare events. 
``Diffusion finally manages to separate the reaction partners.'' \cite{berg1978diffusion}. 
Additionally, these rates can be interpreted as  
\begin{align}
 k_{\mathrm{on}} = k_S\fr{\kappa}{k_S+\kappa} = k_S\phi,
 \label{eq:konrev}
\end{align}
where $\phi=\kappa/(k_S+\kappa)$ can be understood as the fraction of geminate recombinations that 
lead to association or also as the probability of association at the boundary. Then the off rate is 
\begin{align}
 k_{\mathrm{off}} = \mu \fr{k_S}{k_S+\kappa} = \mu(1-\phi),
 \label{eq:koffrev}
\end{align}
where the probability of dissociation at the boundary is
$\big(1-\phi\big)$. The reverse reaction rate constant $k_{\mathrm{off}}$ has the property that it is diffusion controlled if, 
and only if, $k_{\mathrm{on}}$ is diffusion controlled, quite irrespective of the value of $\mu$. ``[A]ny 
description of the reaction process that divides the initial bimolecular event into a 
diffusive association step [$\cdots$] and a subsequent unimolecular transformation is 
logically incorrect.'' \cite{schurr-1970}. This means that in the case of reversible 
bimolecular reactions, we should not treat association and dissociation as independent processes
\footnote{We should note this is not strictly true in the case of particle-based simulations. The collective 
mean behavior of a reaction-diffusion particle-based simulation that models association and dissociation 
independently can have an emergent global coupled behavior. Therefore, particle-based simulations that model 
association and dissociation independently could in theory still reproduce the results established by
\cite{schurr-1970}. As a matter of fact, the model we present validates previous particle-based 
algorithms in the back-reaction boundary condition context from \cite{schurr-1970}}.

For simplicity, all the models presented in this section assumed there is no 
force field. However, all the models have generalizations that 
introduce a force field through a potential; they can all be found in the literature 
cited.

\section{Reaction-diffusion as a discrete stochastic process}\label{sec:reactasstoch}
In this section, we will present the theoretical basis and motivation to model 
reaction-diffusion as a discrete stochastic process. As mentioned in the previous section, 
all the subtle issues that cause confusion among the existing theories of reversible 
reactions with diffusion become trivial when moving to an appropriate discrete time/space 
stochastic description. This will shed some light in the advantages of the discrete 
stochastic models presented in the next section.  It should be noted that similar successful
attempts have been made to write one and two dimensional irreversible Smoluchowski type 
models as discrete stochastic processes 
\cite{torney1983diffusion2,torney1983diffusion,weiss1986overview}. We will break down 
our attention into three different aspects:
\begin{enumerate}
 \item Reactions as stochastic processes and their two descriptions, e.g.,
ensemble distribution and sample trajectories.
 \item Definition of bound/unbound in $A+B\xrightleftharpoons[]{} C$.
 \item Simulations with geminate recombinations.  Gemination is the process described by $\tau_{\kappa}$ in Fig. \ref{fig:FPTs}.
\end{enumerate}
Although these three aspects are closely related to each other, the sequential presentation
illustrates some of the advantages of our discrete stochastic model from slightly different angles.   
Some of the subtle issues that come up in the classical models will be discussed in the 
different sections, and it will be addressed how the discrete stochastic description helps 
solving them. Each of these aspects will be covered in detail in the following subsections.

\subsection{Reactions as stochastic processes and their two descriptions} \label{sec:reactstoch}
Well stirred chemical reactions where there is no spatial component in the equations have 
been successfully modeled with deterministic models based on the Law of Mass Action (LMA). However, for 
cellular biochemical processes inside individual cells, the number of molecules might not be large enough for an accurate continuous 
description. In this case, there is a unifying stochastic mathematical 
framework known as the Delbr{\"u}ck-Gillespie process, whose Kolmogorov forward equation (KFE) has been known as the 
Chemical Master Equation (CME), and whose stochastic trajectories can be computed with the Gillespie, or Doob-Bortz-Kalos-Lebowitz,
algorithm \cite{qian2010cellular,qian2011nonlinear}. Although spatial homogeneity is still assumed in this theory, its depth and richness lie in
Kurtz's theorem, stochastic nonlinear bistability \cite{qian2010chemical},
and stochastic oscillations.  This theory has provided a better understanding of 
biochemical bistability, as shown by a deterministic and 
stochastic comparison of the Schl{\"o}gl model \cite{vellela2009stochastic}, and in the separation 
between the dynamics at short and long time scales.

The Delbr{\"u}ck-Gillespie process is a landmark example, at least in 
chemical reaction dynamics, of the two 
parallel descriptions of a stochastic process. Its CME describes the time-evolution of the probability distribution 
while the Gillespie algorithm describes the stochastic trajectory of 
the system, one reaction at a time,
whose statistical ensemble properties will satisfy such probability distribution. The same dual description is of key 
importance when it comes to producing particle-based numerical simulations. One may know 
(or think to know) the KFE of a diffusion process; however, if we cannot obtain the stochastic 
trajectories from it, it becomes very difficult to produce accurate particle-based numerical simulations.  
In theoretical mathematics, this is the task of constructing a Markov process from its
infinitesimal generator \cite{oksendal2013stochastic}.
This is precisely the case of the continuous reversible diffusion-influenced reactions 
modeled with the back-reaction boundary condition from Eq. \ref{backBC}. The boundary condition 
obscures the underlying stochastic process and consequently complicates any particle-based simulation.
This is the reason why we need a more transparent formulation as the one presented in this 
paper, where the nature of the stochastic process is simple from the beginning.
On the theoretical side, we have also noted that while Kurtz's theorem provides a 
rigorous and satisfying mathematical foundation, in probabilistic terms, 
for the LMA, the theory of hydrodynamic limit as the foundation of nonlinear
partial differential equations of reaction-diffusion type is far from
complete and still only accessible to experts \cite{arnold1980consistency,feng1996hydrodynamic}.

\subsection{Definition of bound/unbound state in $A+B\xrightleftharpoons[]{ \quad } C$} \label{sec:boundun}
Chemical reactions are considered as discrete events.  But at an atomic scale, the very definition 
of a $B$ that is bound ($C$ state) or unbound to
an $A$ molecule is not unequivocal.  This is particularly the case
for diffusion-influenced reaction \cite{qian_j_math_biol}, which can 
yield significantly different quantitative descriptions due to different
laboratory measurements.   We give two such possibilities,
both have been used in experiments that define $C$:
\begin{itemize}
 \item \textbf{Distance}:  FRET 
(F\"{o}rster resonance energy transfer) method directly measures
the distance between two optical markers that are attached on 
$A$ and $B$ respectively.  Therefore, for this type of data, 
we say molecule $B$ is bound to $A$ if the distance between the two molecules is less 
or equal to $\sigma$. We call $r=\sigma$ the absorption or reaction boundary.

\item \textbf{State}: Spectroscopic methods differentiate atomic structures
of a particular chemical group inside a molecule.  If an optically
active chemical group (OACG) in $B$ normally adopts a structure $b$ but a very different 
structure $b^*$ near $A$, then the spectroscopic signals become 
a definition for $B$ versus $C$. Therefore, we say molecule $B$ is bound to $A$ if 
the OACG inside $B$ is in the $b^*$ state; the
molecule is unbound if the OACG is in the $b$ state. 
However, all atomic structures in a molecule 
fluctuate.  In a stochastic setting, the group in $B$ still has a finite 
probability of adopting the $b^*$ structure, while also having a non-zero
probability of being in state $b$ when very near $A$. Note in 
biochemistry, the biological function of a molecule is usually
associated with the state of a chemical group within.
\end{itemize}

In a well-defined bimolecular reaction, the small probabilities 
of a $B$ with its OACG in the $b^*$ state, and a $B$ bound to 
$A$ with its OACG in the $b$ state, are negligibly small.
Therefore, the two definitions above are usually equivalent.

\begin{figure}[h]
  \centering
  \includegraphics[width=0.3\textwidth]{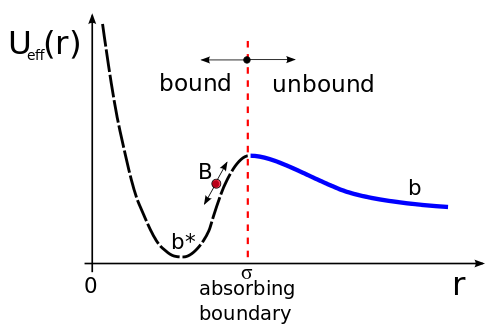}
  \includegraphics[width=0.3\textwidth]{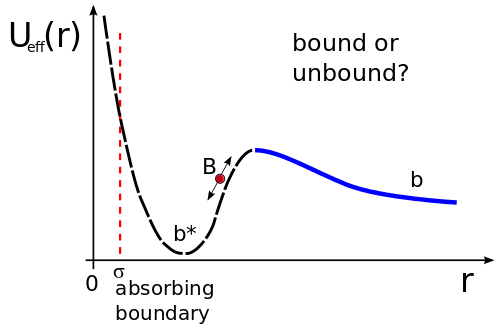} 
  \caption{Plots of the effective interaction potential $U_{eff}(r)=V\left[\left(\fr{r_m}{r}\right)^{12} - 
  2\left(\fr{r_m}{r}\right)^6\right] - \ln{4\pi r^2}$ that molecule 
  $B$ undergoes, where $r$ is the inter-particle distance, $V$
  is the depth of the local minima and $r_m$ the local minima position. 
  The effective potential takes into account a Lennard-Jones potential 
  and the three dimensional geometrical drift terms. The corresponding $b^*$ and $b$ states of the OACG
  are marked in the potential. The left plot (a) shows when the 
  absorbing boundary location matches the expected behavior under the interaction potential. 
  The plot on the right (b) shows an example that the absorbing boundary location can be 
  chosen arbitrarily. In the latter case, it is not clear that the absorbing boundary models 
  the reaction correctly.}
  \label{fig:LJpotential}
\end{figure}

As a simple example, consider a molecule $B$ diffusing in three dimensions with particle $A$
fixed at the origin $r=0$. Assume the intra-molecular potential is described by a Lennard-Jones 
potential. In this case, it is possible to choose the location of the absorption boundary
in the separatrix of the effective potential, so the \emph{state} and \emph{distance} definition match 
each other, as shown in Fig. \ref{fig:LJpotential}a. However, there are many other possibilities. For instance, in Fig. 
\ref{fig:LJpotential}b, we chose the absorption boundary arbitrarily, so it is 
necessary to decide if $B$ is bound to $A$ when it crosses the absorption boundary 
(\emph{distance}) or when it is close to the local minima of $U_{eff}(r)$ (\emph{state}).  
In the latter case, it is not even clear the absorbing boundary can model the reaction accurately, 
so we have to ask the question: does it even make sense to have an absorbing boundary?

The answer is: it depends on the physics one is trying to model. For instance, in the context 
of diffusion under an interaction potential, all irreversible reaction models need an 
absorbing boundary. Without it, there is always a nonzero probability of 
the full reverse reaction to occur. In this case, the choice of the boundary location cannot be completely 
arbitrary; it needs to be chosen based on experimental data or obtained via more fundamental interactions, such as
the Lennard-Jones interaction. One approach that could provide a solution would relate Kramers' 
theory \cite{hanggi1990reaction,kramers1940brownian} applied to the diffusion under a Lennard-Jones 
potential with Smoluchowski's type models. A similar attempt has been studied before for a double well 
potential elsewhere \cite{shoup1982role}.

In the case of reversible reactions, a similar reasoning can be applied, though the concept of absorbing 
boundary needs to be replaced by a reaction boundary. However, it is also possible to model reversible 
reactions without employing a reaction boundary at all. They can be modeled as a diffusion process under 
a bistable effective interaction potential, like the one shown in Figs. \ref{fig:LJpotential}, where one state
corresponds to $B$ being bound and the other to $B$ being unbound. As a matter of fact, any model 
that employs a reaction boundary should be fitted to this more general view of a reversible reaction.
Although there has been some work in this direction \cite{shoup1982role}, it is still not yet fully understood
how to establish the connection between Kramers' theory applied to reversible unimolecular reaction
 \cite{hanggi1990reaction,kramers1940brownian} and the theory of reversible bimolecular reaction mediated by diffusion.

A more complicated example is given in Fig. \ref{fig:KRpotential}a, where the state of the OACG
can oscillate stochastically independent of the distance $r$. In this case, the effective potential is much 
more complicated than just a Lennard-Jones potential. As a matter of fact, it is a potential landscape that can depend on 
more variables than just $r$. In this scenario, the state definition can be extended, at least conceptually, 
to very complicated potential landscapes that could even be stochastic and dynamic. For instance, in Fig. \ref{fig:KRpotential}b, 
the two OACG states are given in terms of two Gibbs free energy potentials $(G)$ as a function of 
the reaction coordinate $q$, like in Marcus theory of electron transfer \cite{marcus1985electron}. In these cases, the 
state description could become completely detached from the distance description. 

Following this analysis, we can say that describing a reversible bimolecular reaction solely through diffusion within
its multivariate potential landscape is a more general approach, which corresponds to the \emph{state} definition. 
Conversely, the models with a reaction boundary are an approximation to this more
general description, which would correspond to the \emph{distance} definition.
This two definitions can be matched if the potential only depends on the inter-particle distance $r$, and we 
choose a distance close to the separatrix of the potential. It should be noted there 
is always the possibility that the potential corresponds to an external 
forcing, or that the absorbing/reacting boundary is actually modeling an absorbing membrane.

These results bring attention to the fact that reaction boundaries are in some sense
``artificial''. From a continuous mathematical point of view, a Brownian particle could 
cross a spatial boundary an infinite amount of times in a finite time, which provides a very 
non-intuitive description of what is physically happening at the reaction boundary and how to 
implement a discrete particle-based simulation. This is also intimately related to the issue of geminate
recombinations, and it is partly the reason why particle-based 
modeling of reaction boundaries is non-trivial. In order to do so, one needs to discretize continuous Brownian 
Motion into a random walk, then somehow provide a discrete model of the reaction boundary, such 
that in the continuous limit the ensemble average over many identical systems satisfies the continuous equation 
and the reaction boundary condition. One can come up with multiple ideas on how to do this, but 
providing one with mathematical rigor is not trivial \cite{burdzy-chen-2008}, especially when the 
reaction boundaries are complex like in Eq. \ref{backBC}.

\subsection{Geminate recombinations}\label{sec:geminate}

Geminate recombinations occur when a particle $B$ that just dissociated from a certain $A$, 
immediately associates again with it. They have been a subject of extensive 
research \cite{agmon1990theory,andrews2004stochastic,khokhlova2012comparison,popov2001three,prustel2012exact}, 
and they are fundamental in providing accurate stochastic reaction-diffusion models and algorithms at cellular and sub-cellular level.
They are also intimately related to the definition of bound and unbound we discussed in the previous subsection.

\begin{figure}[h]
  \centering
  \includegraphics[width=0.3\textwidth]{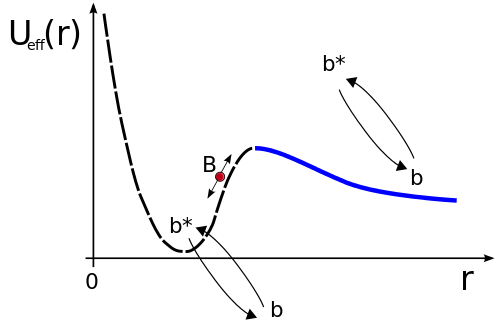}
  \includegraphics[width=0.3\textwidth]{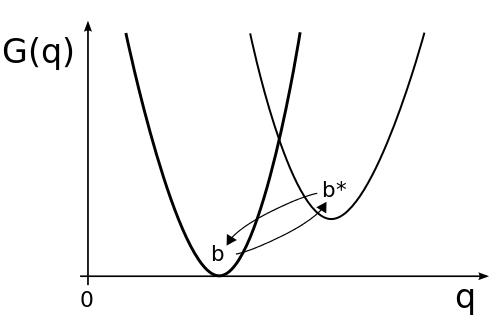}
  \includegraphics[width=0.3\textwidth]{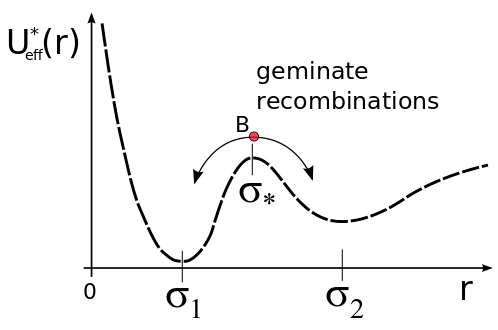} 
  \caption{The left plot (a) is the same than Fig. \ref{fig:LJpotential} with the addition
  of stochastic oscillations of the OACG that are independent of $r$. The center plot (b)
  shows the Gibbs free energy potential for the two possible states of the $B$ molecule as a function of the reaction 
  coordinate. The right plot (c),
  shows how geminate recombinations can be interpreted in the context of diffusion under
  the interaction potential $U_{eff}^*(r)$, 
  where $r$ is the inter-particle distance.
  This effective potential takes into account a double well potential 
  and the three dimensional geometrical drift terms.}
  \label{fig:KRpotential}
\end{figure}

In order to gain a better physical insight, we can observe them in the context of irreversible reactions with partially absorbing
boundaries and Kramers' theory \cite{hanggi1990reaction,kramers1940brownian}. Consider three-dimensional diffusion of a $B$ molecule
under a double well potential $U_{eff}^*(r)$, like shown in Fig. \ref{fig:KRpotential}c. 
Two models will be considered. The first one uses a fully
absorbing boundary at the first well $r=\sigma_1$ and the second one a partially absorbing boundary at the second well $r=\sigma_2$. 
It has been shown that in the steady state the $\kappa$ parameter of the partially absorbing boundary can be calculated using 
Kramers' theory, allowing the two models reaction rate to match each other \cite{shoup1982role}. This result shows that the 
partially absorbing boundary at $r=\sigma_2$ is equivalent to a fully absorbing one at $r=\sigma_1$ \footnote{
The density profile of both approaches is expected to have at least the same shape, although they might integrate to a slightly different 
value due to small differences between the time-dependent rates of the two approaches.}. As we know from 
\cite{berg1978diffusion,burdzy-chen-2008,grebenkov2006,schurr-1970}, partially 
absorbing boundary conditions can be understood as a model for geminate recombinations. The picture just presented favors this view,
since the partially absorbing boundary at $\sigma_2$ can effectively model the back and forth diffusion in the interval 
($\sigma_1$,$\sigma_2$) before the reaction is finally completed. This back and forth diffusion is also represented by the
rebinding diffusion process previously observed in Fig. \ref{fig:FPTs}, and it is what we understand as geminate recombinations.

The previous result states that our partially absorbing models of irreversible reactions do take into
account geminate recombinations. However, there is still an absorbing boundary, so the dissociation process
is not fully modeled. In order to do so, we need to introduce the back-reaction boundary condition at the 
reaction boundary, which is given by a 
partially absorbing boundary condition with the 
addition of the back-reaction term, as in Eq. \ref{backBC}. This model was first 
solved exactly in one dimension and later on in two and three dimensions
\cite{agmon1984diffusion,kim1999exact,prustel2012exact}. However, the solution is not simple, and it's hard to 
grasp some physical intuition out of it. It is also based on a non-local boundary condition, and 
it is not clear, from a rigorous mathematical point of view, what is the corresponding stochastic process 
behind the partial differential equation (PDE) for the probability. This last issue is relevant because it doesn't 
allow us to accurately derive a particle-based stochastic simulation of the trajectories that in average will 
satisfy the probability described by the PDE. Algorithms like eGFRD \cite{takahashi2010spatio} and
FPKMC \cite{donev2010first} are used to simulate reversible reactions using exact solutions for 
reactions between an isolated pair. However, they do not use solutions for the 
back-reaction boundary; they use the solution for the partially absorbing boundary and model 
the dissociation process with an exponential waiting time. It is not obvious nor trivial to show 
that these two approaches are equivalent.

The model presented in the next section bypasses these issues by framing the problem as a discrete Markov 
chain. Its Markovian nature allows simple and accurate particle-based simulations that obey the expected 
statistical behavior given by the KFE. The events occurring at the reaction boundary, including geminate recombinations, are 
intuitive and easily obtained from the jump probabilities.
It also provides a robust and consistent stochastic description that can model reactions using a \emph{distance} 
or \emph{state} approach. In the continuous limit, it recovers the classical models along with the different 
boundary conditions. It also works as a unifying model since it shows other 
approaches to model reversible reactions like eGFRD and FPKMC are consistent with the 
stochastic trajectories of the PDE with the back-reaction boundary condition. It provides further validation for the continuous back-reaction
model for diffusion-influenced reactions since the more intuitive discrete reversible reaction particle-based 
algorithm converges to it. The discrete model's 
robustness and particle-based simulations simplicity lie in its rigorous formulation as a discrete stochastic process 
and its consequent intrinsic relation between the probability distribution dynamics and its individual stochastic trajectories.

\section{The discrete stochastic model for diffusion-influenced reactions}\label{sec:discretemodel}
The main idea is to create an intuitive Markov jump process, that can be interpreted as
a discretization of Eq. \ref{eq:smol2} that conserves probability. The reaction process at the reaction boundaries
is intuitively modeled adding association and/or dissociation jump probabilities. On the continuous limit, 
the continuous equation and the different boundary conditions are recovered, specifically providing 
additional validation for Eq. \ref{eq:backreac_BC}. As it is a discrete time and space stochastic model (Markov chain), 
it is easy to implement a numerical simulation for the probability mass function as well as a particle based simulation for 
the individual stochastic trajectories of the reaction process. We will begin by deriving the framework of the model with a 
radial random walk, and we will add complexity as we recover the different classical diffusion-influenced
reaction models. It should be noted that in all the next models we can take the 
limit $dt\rightarrow0$ to obtain a master equation and use a variant of the Gillespie Algorithm 
to solve it numerically. However, we believe the discrete time approach is educational since it provides a clear
connection between the parameters in diffusion-influenced theory and their probabilistic meaning in particle-based simulations,
see Eqs. \ref{eq:Dkdisc} in Appendix \ref{sec:appB}. 

\subsection{Radial random walk with spherical symmetry}\label{sec:randwalk}
We would like to construct a random walk with spherical symmetry that recovers the Brownian motion diffusion equation in the continuous limit.
We will start by considering a particle following a random walk in spherical coordinates. We will only be interested in the jumps between
different spherical shells in the $r$ direction separated a distance $\delta_r$ and not along the angular directions. 
If the particle is in shell $i$ with radius $r_i = \sigma + i \delta r$ and $\sigma$ a constant, the probabilities to 
jump to the smaller and bigger shells are $p_i$ and $q_i$ respectively. The process is partially illustrated in Fig. \ref{fig:shells}. 

We can write this process as a discrete time Markov chain. Let the position in the radial direction be denoted by $\mathcal{R}_{t}$, 
a random variable for every given $t$ constituting a discrete state and time stochastic process. We will 
call $\pi^{t}_i = \Pr[\mathcal{R}_{t}=r_i]$ the probability of being at spherical shell $i$ at time $t$. The 
state of the whole system at time $t$ is given by the vector of all 
states $\boldsymbol\pi^t =\left[\pi^{t}_0, \pi^{t}_1, \cdots, \pi^{t}_i, \cdots\right]$. The dynamics of our 
random walk are given in terms of the stochastic matrix $\mathbb{P}$ and the Kolmogorov forward equation,
\begin{align}
 \boldsymbol\pi^{t+1} = \boldsymbol\pi^t \mathbb{P}.
 \label{KFE}
\end{align}
Note the stochastic matrix should depend on the probabilities $p_i$ and $q_i$. The resulting stochastic matrix is
{\footnotesize
\begin{align}
 \mathbb{P} = \left[ \begin{matrix}
 1-q_0         & q_0          & 0           & \cdots      &        & \cdots \\ 
 p_1           & 1- (p_1+q_1) & q_1         & 0           &        &        \\
 \vdots        &              & \ddots      &             &        &        \\
               & 0            & p_i         & 1-(p_i+q_i) & q_i    & \cdots  \\      
 \vdots        &              &             & \vdots      & \ddots &        \end{matrix} \right]
 \label{eq:stochmatrix}
\end{align}
}
Note the rows sum to one as expected. For practical purposes, the matrix can be truncated using a 
finite number of shells $N$. In all theses models, we will assume no-flux boundary conditions in the outermost shell, unless stated otherwise.
% Unless stated otherwise, it will be assumed it is placed far away enough to observe relevant physical effects ($N \sim 100$). 
In order to recover Brownian diffusion, we need to adjust the jump probabilities in the random walk in spherical coordinates with 
\begin{align}
\begin{split}
p_i=\delta t\left(\fr{D}{\delta r^2} - \fr{D}{r_{i-1}\delta r}\right), \ \ \ \
q_i=\delta t\left(\fr{D}{\delta r^2} + \fr{D}{r_{i+1}\delta r}\right), 
\end{split}
\label{probdiff}
\end{align}
where $D$ is the constant diffusion coefficient. Note the probability of staying in the shell $i$ given by $1-(p_i+q_i)$ grows 
approaching $1-2\delta t D/\delta r^2$ as $r_i$ is increased. Using these values we can rewrite the $i^{\mathrm{th}}$ equation 
of (\ref{KFE}) as,
\begin{align*}
 \pi_i^{t+1} = \pi_i^{t} + \delta t D \left[ \fr{\pi_{i-1}^{t} -2\pi_i^{t} +\pi_{i+1}^{t}}{\delta r^2} \right] - 
               \delta t \fr{2D}{r_i}\left[\fr{\pi_{i+1}^{t} - \pi_{i-1}^{t}}{2\delta r}\right] + 
                \delta t \fr{D}{\delta r}\left[\fr{\pi_i^{t}}{r_i - \delta r} - \fr{\pi_i^{t}}{r_i + \delta r}\right].
\end{align*}
We would like to recover a continuous equation for the probability. The discrete probability $\pi_i^t$
is related to the continuous probability distribution function $\Pi(r_i,t)$ by $\pi_i^t = \Pi(r_i,t) \delta r$, 
so we need to divide $\pi_k^t$ by $\delta r$ for all $k$ first. However this doesn't make any difference in structure of 
the equation, so now we can take the limit as $\delta_t \rightarrow 0 $ and $\delta_r \rightarrow 0$ following standard finite
difference theory \cite{randysbbook}, which yields second order accuracy in space and first order in time. We obtain,
\begin{align}
 \fr{\p \Pi(r,t)}{\p t} = D\fr{\p^2 \Pi(r,t)}{\p r^2} - \fr{2D}{r}\fr{\p \Pi(r,t)}{\p r} + \fr{2D}{r^2}\Pi(r,t),
 \label{eq:contlim}
\end{align}
where $\Pi(r,t)$ is a probability distirbution function in $r$. This equation can be written in the form of a Fokker Planck equation with drift,
\begin{align}
 \fr{\p \Pi(r,t)}{\p t} = D\fr{\p^2 \Pi(r,t)}{\p r^2} - \fr{\p}{\p r}\left( \fr{2D}{r} \Pi(r,t)\right).
 \label{eq:FPEdr}
\end{align}
The probability of being a distance $r$ from the origin at time $t$ is $\Pr\{r < \tilde{\mathcal{R}}_{t} \le r+ \delta r\} = \Pi(r,t) dr$, where 
$\tilde{\mathcal{R}}_{t}$ is now a stochastic process with continuous state and time. However, this is the probability of being at any point in
the sphere with radius $r$, so we cannot yet compare it with the Smoluchowski diffusion equation. In order to do so, we need 
the equation for the probability of being at any point in space given by $f(r,\theta,\phi,t)r^2 \sin(\theta) dr d\theta d \phi$. 
Integrating this equation in the angular coordinates yield the probability $\Pi(r,t)$. As we have spherical symmetry in $f$, this yields 
\begin{align}
 \Pi(r,t)dr = 4\pi r^2 f(r,t)dr.
 \label{eq:BMtrans}
\end{align}
Substituting this result into Eq. \ref{eq:FPEdr} and doing some algebra, we recover the Smoluchowski equation (Eq. \ref{eq:smol}) for 
constant $D$ as expected,
\begin{align}
 \fr{\p f(r,t)}{\p t} = \fr{D}{r^2} \fr{\p}{\p r} \left(r^2 \fr{\p f(r,t)}{\p r}\right) = \nabla \cdot[D \nabla f(r,t)],
 \label{eq:heat}
\end{align}
where $\nabla$ operates in spherical coordinates with symmetry in the polar and azimuthal angle; therefore, the discrete time Markov 
chain models the discrete analog to overdamped Brownian motion in spherical coordinates. However, we haven't yet discussed the 
continuous limit in the discrete boundary of the innermost shell $r_0$. We can also write the first equation of (\ref{KFE}) as,
\begin{align*}
 \pi_0&^{t+1} = \pi_0^{t} + \delta t D \left[ \fr{\pi_{1}^{t} -\pi_0^{t}}{\delta r^2} \right] - 
      \delta t \left[\fr{D}{r_0}\fr{\pi_{1}^{t}}{\delta r} + \fr{D}{\delta r} \left(\fr{\pi_0^{t}}{r_0 + \delta r}\right)\right]. \\[2mm]
\end{align*}
In order to find what boundary condition this discretization satisfies, we will 
follow standard finite difference techniques \cite{randysbbook}. We will introduce the ghost cell $\pi_{-1}^{t}$, so 
we can rewrite this equation as,
\begin{align*} 
   \pi_0&^{t+1} = \pi_0^{t} + \delta t D \left[ \fr{\pi_{1}^{t} -2\pi_0^{t} +\pi_{-1}^{t}}{\delta r^2} \right] - 
                \delta t \fr{2D}{r_0}\left[\fr{\pi_{1}^{t} - \pi_{-1}^{t}}{2\delta r}\right] + 
                 \delta t \fr{D}{\delta r}\left[\fr{\pi_{0}^{t}}{r_0 - \delta r} - \fr{\pi_0^{t}}{r_0 + \delta r}\right],
\end{align*}
where
\begin{align}
\pi_{-1}^t = \pi_0^t\fr{\left(\fr{1}{\delta r}-\fr{1}{r_{-1}}\right)}{\left(\fr{1}{r_0} + \fr{1}{\delta r}\right)}.
\label{eq:refbc}
\end{align}
Dividing by $\delta r$ and taking the limit  $\delta_t \rightarrow 0$, $\delta_r \rightarrow 0$, we again recover 
Eq. \ref{eq:contlim} with second order accuracy in space and Eq. \ref{eq:refbc} becomes the zero flux condition 
for the Fokker-Planck equation (see Appendix \ref{app:secord}),
\begin{align*}
\left. \fr{\p \Pi(r,t)}{\p r} \right|_{r=r_0}= \fr{2\Pi(r_0,t)}{r_0}.
\end{align*}
Using this result and employing Eq. \ref{eq:BMtrans}, the reflective boundary condition for Eq. \ref{eq:heat} 
can be rewritten as $\left. \fr{\p f(r,t)}{\p r} \right|_{r=r_0} = 0$, as expected. Note we assumed
$r_0 - \delta r > 0$. Unless stated otherwise, in this and all of the 
subsequent models we will also employ a zero flux boundary on the outermost spherical shell 
$r=r_{max}$. As we have an irreducible and aperiodic Markov chain with a finite number of states, 
we know it has a steady state. Moreover, as the flux is zero on both boundaries, the detailed balance 
condition must be satisfied everywhere. Detailed balance will be mentioned in more detail on Section \ref{sec:ur}. In order to recover 
the Smoluchowski diffusion equation (Eq. \ref{eq:smol}) in the whole space, we can take the limit 
as $r_0\rightarrow 0$ and $r_{max}\rightarrow \infty$. 

It should be noted that the limit $\delta_t \rightarrow 0$, $\delta_r \rightarrow 0$ will only guarantee our discretization is 
consistent with the continuous model. In order for the method to be convergent in the finite difference sense, we also need to 
satisfy a CFL condition that we can obtain
through stability analysis \cite{randysbbook}. However, this discretization is also a Markov chain, so we can also obtain a stability condition
by making sure that all the rows of matrix \ref{eq:stochmatrix} sum to one and that all the entries are in the interval $[0,1]$. This 
analysis yields the condition that $\fr{\Delta t}{\Delta r^2}<\fr{1}{2D}$, which happens to be the same as the CFL condition for the one-dimensional 
diffusion equation. When we add more complexity in the next sections, like boundary reaction terms and potentials, we need to be careful 
that our matrix still satisfies these Markov conditions.

On the discrete model, we can also write the diffusion coefficient in terms of the jump probabilities by 
summing up Eqs. \ref{probdiff} to yield Einstein's relation,
\begin{align}
 D = \fr{\delta r^2}{2 \delta t} (p_{i+1} + q_{i-1}).
 \label{eq:einrel}
\end{align}
This is a more general expression than the obtained in random walks on Cartesian coordinates. On a Cartesian random walk, 
if the probability of staying in the same spot is zero, we obtain $p_{i+1} + q_{i-1}=1,1/2,1/3$, in one, two and three 
dimensions respectively, recovering the well known expression $D=\delta r^2/(2n \delta t)$ with $n$ the number of dimensions. 

In the next section, we will use this spherical random walk to construct a discrete model for irreversible diffusion-influenced reactions.  

\subsection{Discrete model for irreversible diffusion-influenced reactions}\label{sec:irrRD}
The random walk derived in the previous model does not yet include any reaction. In order to study the reaction $A+B\rightarrow C$, 
consider again a particle $B$ diffusing under the spherical symmetric random walk in the previous section. In addition, there will be 
a particle $A$ fixed at the origin. If we define the shell $r_0=\sigma$ as the reaction boundary or the binding radius, then we can 
incorporate the probability of a reaction $P_b=\tilde{\kappa}(r)\delta t$. An illustrated description of the process is 
shown in Fig. (\ref{fig:shells}), and a detailed one-dimensional symmetric version of this model is presented in 
Appendix \ref{sec:appB}. The stochastic matrix will now look like 

{\footnotesize
\begin{align}
 \mathbb{P} = 
 \left[ \begin{matrix}
 1-(q_0+ P_b)       & q_0                        & 0           & \cdots      &         \\ 
 p_1                                & 1- (p_1+q_1)               & q_1         & 0           &         \\
 \vdots                             &                            & \ddots      &             &         \\
                                    & 0                          & p_i         & 1-(p_i+q_i) & q_i      \\      
 \vdots                             &                            &             & \vdots      & \ddots  \end{matrix} \right].
 \label{reactmat}
\end{align}
}
Note the first row doesn't sum up to one, since there is a probability of being absorbed, $P_b$. In order to recover 
the total probability, we need to sum this probability to the first row of the stochastic matrix. Furthermore, the 
association rate $\tilde{\kappa}(r)$ will scale depending on where we chose our reaction boundary to be. A physically 
reasonable assumption is that the rate $\tilde{\kappa}(r)$ scales inversely to the infinitesimal volume of the reaction 
spherical shell, $\tilde{\kappa}(r) = \kappa/(4\pi r^2 \delta_r)$, where $\kappa$ is the constant rate in the boundary 
condition of Eq. \ref{backBC} and has units of volume over time. The bigger the shell, the smaller we 
need $\tilde{\kappa}(r)$ to be in order to keep the model consistent. Furthermore, note the 
probability of being absorbed at shell $r_0=\sigma$ is $P_b = \tilde{\kappa}(r_0)\delta t$, so we can solve 
for $\kappa$ to obtain $\kappa = 4\pi \sigma^2 P_b \delta r / \delta t$. This equation, along with Eq. \ref{eq:einrel}, 
provide the diffusion-influenced theory parameters $D$ and $\kappa$ in terms of the jump probabilities. As shown in the Appendix \ref{sec:appB},
the probability of being absorbed could be replaced by a more accurate value $P_b=1-e^{-\tilde{\kappa}(r_0)\delta t}$, 
in which case $\kappa = -4\pi \sigma^2  \delta r \log{[1-P_b]}/ \delta t$. However, this is not relevant in the continuous 
limit analysis we will now carry out.

\begin{figure}
\centering
\includegraphics[width=0.47\columnwidth]{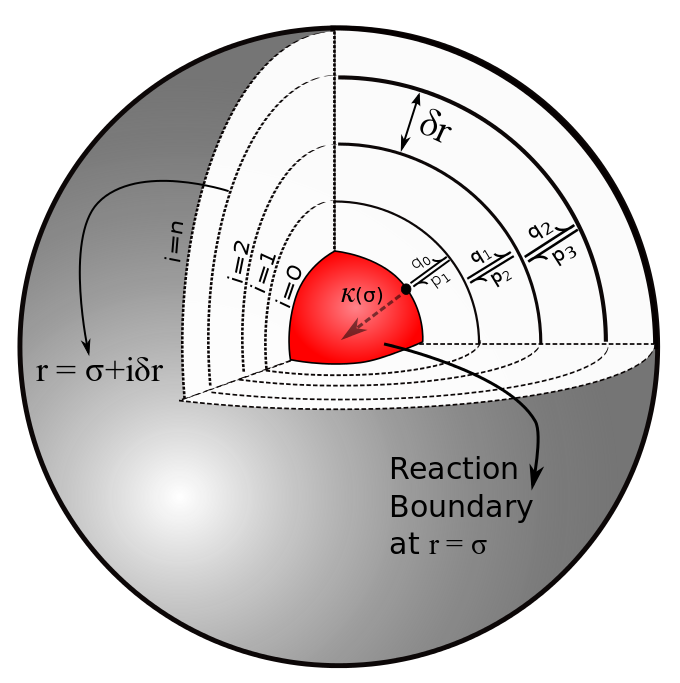}
\caption{Concentric shells for the discrete model for irreversible diffusion-influenced reactions. This figures illustrates a random walk in
spherical coordinates and the reaction occurring at the reaction boundary at shell $i=0$.}
\label{fig:shells}
\end{figure}

Employing this stochastic matrix on the system (\ref{KFE}) yields the same result as in the spherical random walk except for 
the first equation in the boundary $r_0$ which yields,
\begin{align*}
 \pi_0&^{t+1} = \pi_0^{t} + \fr{\delta t D}{\delta r ^2} \left[ \pi_{1}^{t} -2\pi_0^{t} + 
                \pi_0^{t}\left(1-\fr{\delta r ^2}{D}\fr{\kappa}{4\pi r_0^2 \delta r} \right) \right]  
      -\delta t \left[\fr{D}{r_0}\fr{\pi_{1}^{t}}{\delta r} + \fr{D}{\delta r} \left(\fr{\pi_0^{t}}{r_0 + \delta r}\right)\right]. \\[2mm]
  \Rightarrow \ \ \ 
   \pi_0&^{t+1} = \pi_0^{t} + \delta t D \left[ \fr{\pi_{1}^{t} -2\pi_0^{t} +\pi_{-1}^{t}}{\delta r^2} \right] - 
                \delta t \fr{2D}{r_0}\left[\fr{\pi_{1}^{t} - \pi_{-1}^{t}}{2\delta r}\right] + 
                 \delta t \fr{D}{\delta r}\left[\fr{\pi_{0}^{t}}{r_0 - \delta r} - \fr{\pi_0^{t}}{r_0 + \delta r}\right],
\end{align*}
where $\pi_{-1}^t$ satisfies
\begin{align}
\pi_0^t - \fr{\delta r \kappa}{4\pi Dr_0^2}\pi_0^t = \pi_{-1}^t + \delta r\fr{\pi_0^t }{r_{-1}} + \delta r\fr{\pi_{-1}^t }{r_{0}}
\label{eq:CKdisbc}
\end{align}
Dividing by $\delta r$ and taking the limit as $\delta_t \rightarrow 0$, $\delta_r \rightarrow 0$ again satisfies the 
Fokker-Planck equation (Eq. \ref{eq:FPEdr}) with second order accuracy in space; however, Eq. \ref{eq:CKdisbc} now becomes
\begin{align*}
 \left. \fr{\p \Pi(r,t)}{\p r}\right|_{r=\sigma} = \fr{\kappa}{4\pi\sigma^2D}\Pi(\sigma,t) + \fr{2\Pi(\sigma,t)}{\sigma}
\end{align*}
with first order accuracy, where $\Pi(r,t)$ is the continuous analog of $\pi_i^{t}$. Employing once again the change of 
variables in Eq. \ref{eq:BMtrans}, we recover the
well known boundary condition of Collins and Kimball (see Eq. \ref{colBC}),
\begin{gather*}
  4 \pi \sigma^2 D \left. \fr{\p f(r,t)}{\p r} \right|_{r=\sigma} = \kappa f(\sigma,t).
\end{gather*}

We would like to provide a comparison between Collins and Kimball solution (Eq. \ref{eq:cksol}) and our Markov approach;
however, our equations model the dynamics for the probability mass function and Collins and Kimball models the concentration
gradient. A possible probabilistic interpretation of 
Collins and Kimball model is given by the Green,s function for an isolated
pair, like in the GFRD approach \cite{sokolowski2010green,van2005simulating,van2005green}; however, the steady state will
yield zero. The main difficulty in providing a better probabilistic interpretation is the far-field boundary condition
with constant value at infinity. In order to address this issue, we will take an alternate approach to compare the steady state solution. 
We will solve the Fokker-Planck equation (Eq. \ref{eq:smol}) with the boundary conditions,
\begin{align}
 \left. 4 \pi \sigma^2 D \fr{\p f(r,t)}{\p r} \right|_{r=\sigma} = \left. 4 \pi R^2 D \fr{\p f(r,t)}{\p r} \right|_{r=R} = \kappa f(\sigma,t),
 \label{CKperiodic}
\end{align}
and $\int_{\sigma}^R 4\pi r^2 f(r,t) dr = 1$. These conditions mean that the probability
flux at $r=\sigma$ is the same as the flux at $r=R$, which is, from a probabilistic point of view, a periodic boundary
condition. The steady state solution is exactly of the same form as Eq. \ref{eq:cksol}, but with a fixed constant $c_0=A_0$,
\begin{eqnarray}
 f_{ss}(r) &=& A_0\left[1-\fr{\kappa \sigma}{4\pi D \sigma + \kappa} \left(\fr{1}{r}\right)\right], \label{eq:cksolper} \\[2mm]
  A_0 &=& \left[ 4\pi\left(\fr{R^3 -\sigma^3}{3}\right) - \fr{4\pi \sigma \kappa}{4\pi\sigma D + \kappa} 
  \left(\fr{R^2-\sigma^2}{2}\right)\right]^{-1} . \label{eq:A0c0}
  \label{eq:ck}
\end{eqnarray}
This result provides a mathematical connection with the original gradient concentration approach and the probability approach. 
As a matter of fact, from a mathematical point of view, the boundary condition in Eq. \ref{CKperiodic} is also 
satisfied in the original Collins \& Kimball formulation at steady state. In the probabilistic interpretation, the free parameter $A_0$
will give the normalization constant for the probability. In this case, we fixed it so the probability integrates to one. Nonetheless, we 
could have chosen it to integrate to any other value between $0$ and $1$. For instance, if we want the probability to integrate to $0.7$,
we can obtain the corresponding value of $A_0$. In the concentration interpretation and scaled accordingly, this value will correspond to 
the concentration of the material/chemical bath in the far-field in order to get $30\%$ of absorbtion before reaching the steady state.
For our current comparison and without loss of generality, we chose it so the probability integrates to one. Although this might seem odd given
we have an absorbing boundary condition, the far-field boundary condition is a source compensates for the absorption.

\begin{figure}[t!]
    \centering
%    \begin{subfigure}[t]{0.4\textwidth}
        \centering
        \includegraphics[width=0.47\columnwidth]{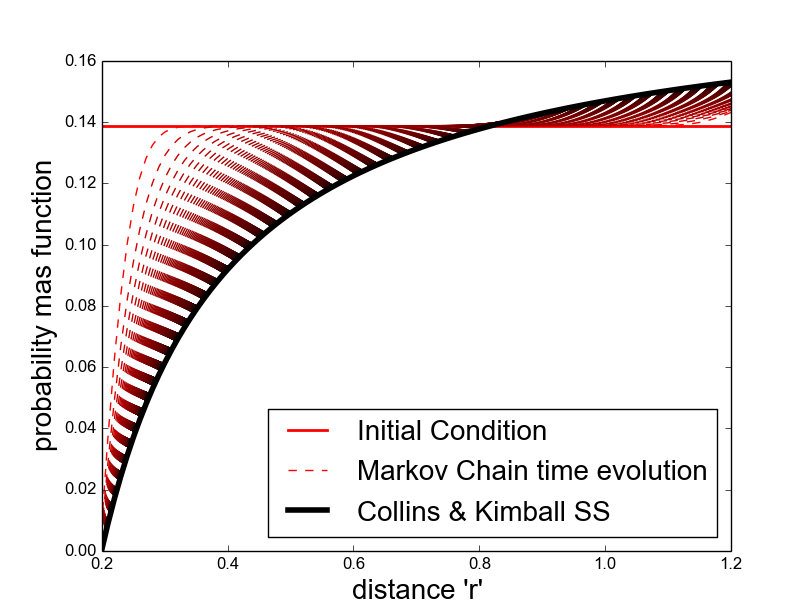} 
%         \caption{}
%    \end{subfigure}
%    \begin{subfigure}[t]{0.4\textwidth}
%         \centering
        \includegraphics[width=0.47\columnwidth]{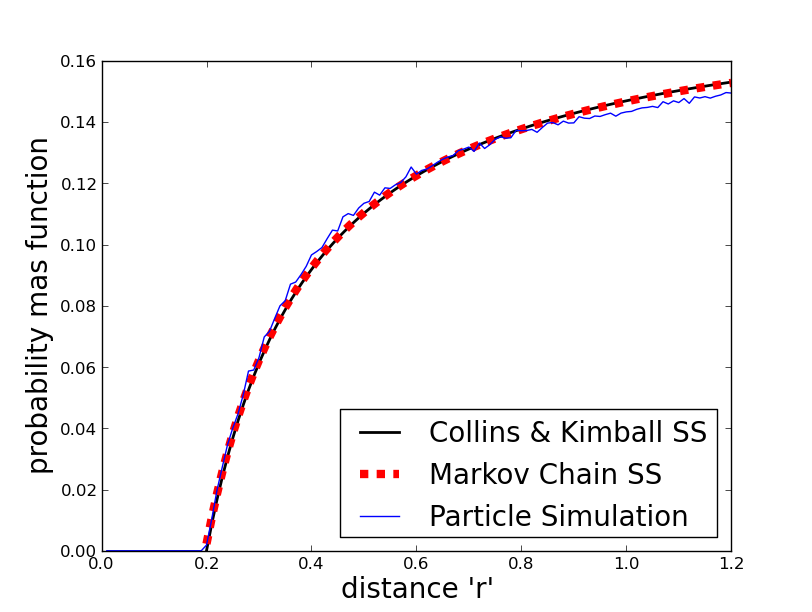} 
%         \caption{}
%    \end{subfigure}
    \caption{Model verification: (a) The steady state of Collins and Kimball's solution of Eq. \ref{eq:cksol} (or Eq. \ref{eq:cksolper})
        with $c_0=A_0$ from Eq. \ref{eq:A0c0} is plotted with a thick black line. The initial condition (uniform distribution) 
        for the discrete Markov model 
        is plotted as an horizontal red line. The dashed lines represent the convergence in time to the steady state of 
        the Markov chain from $t=0$ to $t=1$ taken every 100 time steps; the darker
        lines correspond to longer times. (b) The steady state of the periodic solution to Collins and Kimball (Eq. \ref{eq:cksolper})
        is compared to the Markov chain steady state and to a particle based simulation with 3E6 particles and 1E4 time 
        iterations. The parameters used were: $\delta r = 0.01$, $\delta t = 0.0001$, $D = 0.1$, 
        $\tilde{\kappa}(\sigma) = 4000.0$, $\sigma = 0.2$, and 100 shells for the discrete models.}
        \label{fig:CK_comp}
\end{figure}

In order to do the comparison, we modify the stochastic matrix (\ref{reactmat}) in our Markov model to model this periodicity by adding a 
$\tilde{\kappa}(\sigma)\delta t$ term in the last column and first row of the truncated stochastic matrix. We can compute the probability mass 
function from the Markov chain at time $t=n\delta t$ with $\boldsymbol\pi^{n} = \boldsymbol\pi^0 \mathbb{P}^n$ until reaching the steady state
$\boldsymbol\pi^{ss} = \boldsymbol\pi^{ss} \mathbb{P}$, and we can also do a particle based simulation using the jump
probabilities. A comparison of the three approaches for uniform initial distributions is shown in Figure \ref{fig:CK_comp}. 
The quantity plotted in this and every other figure in this section is $\pi_i^t/(4\pi r_i^2 \delta r)$, which uses the 
first order equation that the probability between the point-continuous and spherical-discrete setting is $\pi_i^t=4\pi r_i^2 f(r_i,t)\delta r$.
Note the agreement between the ensemble behavior of stochastic trajectories and the probability mass function is expected from 
the stochastic theory. This stochastic formulation of our model is what allows trivial particle-based simulations that 
accurately match the expected statistical behavior given by probability mass function dynamics. It should also be noted
this is an open system in non-equilibrium steady state (NESS) driven by the flux from the sink and source at the inner and outer boundary, 
so it will not satisfy detailed balance \cite{qian2007phosphorylation}.

In order to provide an even more complete connection to the meaning of the Collins and Kimball rate in the discrete model,
consider a particle in the shell in the reaction boundary $r_0=\sigma$. The particle has three possible movements: it can diffuse to level $i=1$
with probability $q_0\approx \delta t D/\delta r^2$ (first order); it can react with probability 
$\tilde{\kappa}(\sigma) \delta_t$; or it can diffuse along the spherical shell with probability $\xi$. Lets assume that if the particle 
diffuses out to shell $n=1$ then the particle is fully dissociated 
without any geminate recombinations; therefore, the probability of reaction while diffusing in the shell $r_0$ is given by
\begin{align*}
 \phi = \tilde{\kappa}(\sigma) \delta_t \sum_{n=0}^{\infty} \xi^n = \fr{\tilde{\kappa}(\sigma) \delta_t}{1-\xi} = 
    \fr{\tilde{\kappa}(\sigma) \delta_t}{\tilde{\kappa}(\sigma)\delta_t + \delta t D/\delta r^2} 
\end{align*}
where we used that $1 = \xi + \tilde{\kappa}(\sigma)\delta_t + q_0 $. As we already obtained that $\tilde{\kappa}(r) = \kappa/(4\pi r^2 \delta_r)$, 
we can substitute $\tilde{\kappa}(\sigma)$, which yields
\begin{align*}
 \phi = \fr{\kappa}{4\pi \sigma^2 D/\delta_r + \kappa}.
\end{align*}
This is the discrete approximation of the fraction of reactions due to geminations of the Collins and 
Kimball reaction rate \cite{andrews2004stochastic,berg1978diffusion,schurr-1970}. Its continuous counterpart is shown in Eqs. 
\ref{eq:konrev} and \ref{eq:koffrev}. Note we do not expect convergence as $\delta r \rightarrow 0$, since this approximate 
derivation will no longer make sense from a continuous perspective. Nonetheless, it provides a clear physical picture of how 
the gemination process work.

\subsection{Discrete model for reversible reaction-diffusion}\label{sec:revRD}
In this section, we will extend the previous model to deal with reversible reactions. We will begin by adding one more state $\pi^t_{b}$ at 
the beginning of the state vector: $\boldsymbol{\pi}^t =\left[\pi^t_{b},\pi^{t}_0, \pi^{t}_1, \cdots, \pi^{t}_i, \cdots\right]$. This new 
state means the probability of the $B$ molecule to be bound to $A$. The stochastic matrix will naturally require an additional first row 
and column
{\footnotesize
\begin{align}
 \mathbb{P} = 
 \left[ \begin{matrix}
 1-P_u               & P_u                       & 0                          & 0           & \cdots             \\
 P_b          & 1-(q_0+ P_b)       & q_0                        & 0           & \cdots              \\ 
 0                            & p_1                                & 1- (p_1+q_1)               & q_1         & \cdots                  \\
 \vdots                       &                                    &                            & \ddots      &                       
 \end{matrix} \right],
 \label{reactmat_rev}
\end{align}
}
where $P_b=\tilde{\kappa}(r)\delta t$ and $P_u=\tilde{\mu}(r_0) \delta t$ is the probability for dissociation to occur at the corresponding shell.
The dissociation parameter $\tilde{\mu}(r)$ is trivially related to the constant dissociation rate $\mu$ in 
Eq. \ref{backBC} by $\tilde{\mu}(r)=\mu$ with units of time$^{-1}$. Once again $P_b$ and $P_u$ could be replaced with more accurate 
exponential expressions, see Appendix \ref{sec:appB}.

\begin{figure}[t!]
    \centering
%    \begin{subfigure}[t]{0.4\textwidth}
        \centering
        \includegraphics[width=0.47\columnwidth]{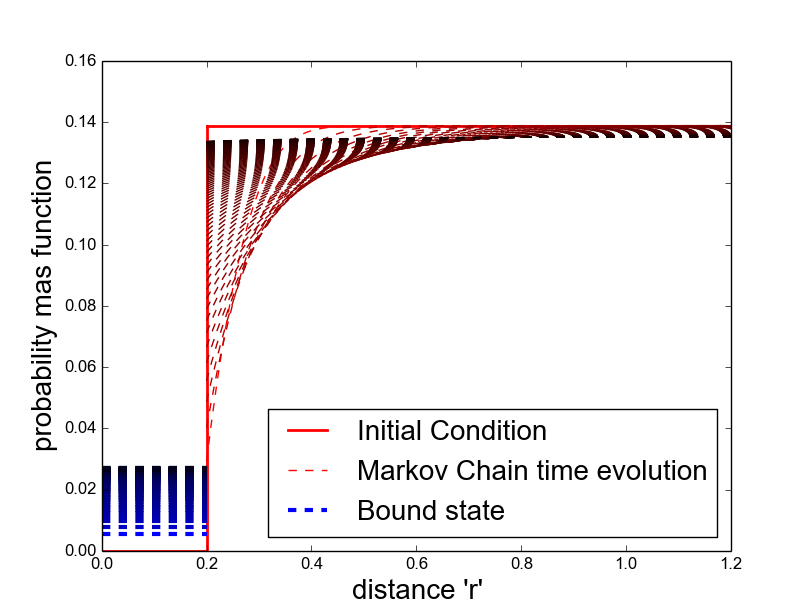} 
%         \caption{}
%    \end{subfigure}
%    \begin{subfigure}[t]{0.4\textwidth}
%         \centering
        \includegraphics[width=0.47\columnwidth]{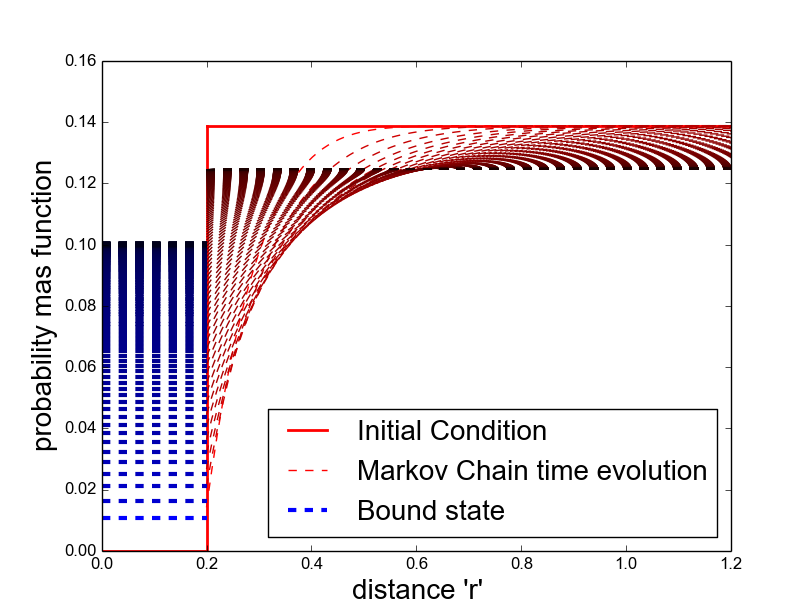} 
%         \caption{}
%    \end{subfigure}
    \caption{Convergence of the discrete model to steady state for the reversible case for two sets of parameters. 
        The initial condition is a uniform distribution in the unbound state and zero on the bound state, 
        and it is plotted as a red line. The dashed lines represent the convergence in time to the steady state of 
        the Markov chain from $t=0$ to $t=1$ taken every 400 and 1000 time steps respectively; the darker
        lines correspond to longer times. (a) Parameters used were: $\delta r = 0.01$, $\delta t = 0.0001$, $D = 0.1$, 
        $\tilde{\kappa}(\sigma) = 8000.0$, $\mu = 200$ $\sigma = 0.2$, and 100 shells. The bound state $\pi_b$ value 
        at different time steps is shown as an horizontal line from $r=[0,\sigma]$ (b) Same parameters 
        with rates exaggerated to better show the convergence to the steady state.}
        \label{fig:rev_conv}
\end{figure}

Note that now all the rows of the stochastic matrix sum to zero since we are taking into account the particles that have already reacted. 
Also note $\pi^t_b$ is the probability of $B$ being bound to $A$, so it can be written as $\pi^t_b = [1 - S_d(t)]$, where 
$S_d(t) = \sum_{i=0}^{N} \pi_i^t$ is the probability of being unbound at time $t$. The first equation now yields,
\begin{align*}
 \pi_{b}^{t+1} = \pi_{b}^{t}\left[1 - \tilde{\mu}(r_0) \delta t\right] + \pi_0^t \tilde{\kappa}(r_0)\delta t.
\end{align*}
Dividing by $\delta r$ and taking the limit, we can arrange the left hand side to yield 
$\Pi_b(t)\approx(\pi_{b}^{t+1} - \pi_{b}^{t})/\delta t$. On the right hand side it might seem this equation might not 
converge, since $\tilde{\kappa}(r_0)$ scale as $1/ \delta r$. However, using Eq. \ref{eq:BMtrans} we know 
$\pi_0^t = 4 \pi r^2 f(r_0,t) \delta r$ at first order, so in the limit we immediately obtain,
\begin{align}
 \fr{d f_b(t)}{d t} = \kappa f(r_0,t) - \mu f_b(t),
 \label{eq:boundReact}
\end{align}
where $f_b(t) = \Pi_b(t)$, so $\mu$ has units of time$^{-1}$ and $\kappa$ of volume over time as expected. 
This equation only involves reaction, since we didn't allow any diffusion to happen when $B$ is in
the bound state. The second equation will yield even more interesting results, the equation is
\begin{align*}
 \pi_0^{t+1} = \pi_0^{t} + \fr{\delta t D}{\delta r ^2} \left[ \pi_{1}^{t} -2\pi_0^{t} + 
                \pi_0^{t}\left(1-\fr{\delta r ^2}{D}\fr{\kappa}{4\pi r_0^2 \delta r} \right) \right]   
      -\delta t \left[\fr{D}{r_0}\fr{\pi_{1}^{t}}{\delta r} + \fr{D}{\delta r} \left(\fr{\pi_0^{t}}{r_0 + \delta r}\right)\right]
      + \delta t \mu \pi_b^t. \\[2mm]
  \Rightarrow \ \ \ 
   \pi_0^{t+1} = \pi_0^{t} + \delta t D \left[ \fr{\pi_{1}^{t} -2\pi_0^{t} +\pi_{-1}^{t}}{\delta r^2} \right] - 
                \delta t \fr{2D}{r_0}\left[\fr{\pi_{1}^{t} - \pi_{-1}^{t}}{2\delta r}\right] + 
                 \delta t \fr{D}{\delta r}\left[\fr{\pi_{0}^{t}}{r_0 - \delta r} - \fr{\pi_0^{t}}{r_0 + \delta r}\right],
\end{align*}
where $\pi_{-1}^t$ satisfies
\begin{align}
\pi_0^t - \fr{\delta r \kappa}{4\pi Dr_0^2}\pi_0^t + \mu\fr{\delta r^2}{D} \pi_b^t = 
\pi_{-1}^t + \delta r\fr{\pi_0^t }{r_{-1}} + \delta r\fr{\pi_{-1}^t }{r_{0}}.
\label{eq:KSdisbc}
\end{align}
Dividing by $\delta r$ and taking the limit as $\delta_t \rightarrow 0$, $\delta_r \rightarrow 0$ again satisfies 
the Fokker-Planck equation (Eq. \ref{eq:FPEdr}). For Eq. \ref{eq:KSdisbc},
we use again Eq. \ref{eq:BMtrans} and the fact that $\pi_i^t = 4\pi r^2 f(r_i,t)\delta r$ at first order to obtain 
\begin{gather*}
  4 \pi \sigma^2 D \left. \fr{\p f(r,t)}{\p r} \right|_{r=\sigma} = \kappa f(\sigma,t) - \mu f_b(t),
\end{gather*}
where $f_b(t)=\Pi_b(t)$ and $\sigma=r_0$. Now note that $f_b(t) = [1-S(t)]$, where $S(t)=\int_\sigma^\infty 4 \pi r^2f(r,t) dr$ is the probability 
of being unbound, i.e. the continuous version of $S_d(t)$, so we can write
\begin{align}
  4 \pi \sigma^2 D \left. \fr{\p f(r,t)}{\p r} \right|_{r=\sigma} = \kappa f(\sigma,t) - \mu [1-S(t)],
  \label{eq:backreac_BC}
\end{align}
This is the back-reaction boundary condition (Eq. \ref{backBC}) found in the exact solution for the reversible 
reaction \cite{agmon1984diffusion,kim1999exact,prustel2012exact}. Note this boundary condition couples the Fokker-Planck 
equation (Eq. \ref{eq:FPEdr}) with the Eq. \ref{eq:boundReact}. The probability of being unbound can be represented in terms 
of the integral in Eq. \ref{eq:survprob}, which yields a non-local boundary condition and sheds light in the fact that 
some non-local boundary conditions might be only a condensed technique to write complicated coupled systems. The time 
evolution of the probability mass function given by our Markov approach is shown for two set of parameters in Figure \ref{fig:rev_conv}. 
The initial condition is given as uniform for the unbound state and zero for the bound one. The probability is conserved following, 
\begin{align*}
 \pi_b^t + \sum_{i=0}^{N} \pi_i^t = 1, 
\end{align*}
for all times $t$, where $\pi_i^t$ is equal to $4\pi r_i^2 f(r_i,t) \delta r$ at first order, and 
$\pi_b$ is the probability of being bound---its value is 
represented by the height of the blue dashed bar on the plots in Figure \ref{fig:rev_conv}. On both plots we
can observe convergence towards a flat steady state. As we included the bound state
as part of the irreducible and aperiodic Markov chain as well as no-flux boundaires elsewhere, the system is closed 
and has a steady state, so detailed balance must be satisifed. The probability lost in the unbound region is 
balanced by the probability gained in the 
bound one. Since the model is a Markov chain, particle-based simulation of this
discrete stochastic reversible diffusion-influenced model is trivial by employing the 
jump probabilities.

Note the continuous limit we just derived of the discrete model provides a more simple form of the 
model than the original back-reaction continuous model as two coupled partial/ordinary differential equations,
\begin{align}
 \fr{d f_b(t)}{d t} = \kappa f(r_0,t) - \mu f_b(t),  \label{eq:BR_newmodel1}\\
 \fr{\p f(r,t)}{\p t} = \nabla \cdot[D(r) \nabla f(r,t)] \label{eq:BR_newmodel2}
\end{align}
 with the back-reaction boundary condition
\begin{align}
   4 \pi \sigma^2 D \left. \fr{\p f(r,t)}{\p r} \right|_{r=\sigma} = \kappa f(\sigma,t) 
  - \underbrace{\mu f_b(t)}_{\text{$\mu[1-S(t|r_0)]$}}. \label{eq:BR_newmodel3}
\end{align}

Although this equation can be derived from the original model, we recovered it naturally 
from the discrete model limit. As the association and dissociation were intuitively implemented
in the discrete model, its convergence to the original continuous model, provides additional
validation that it is appropriate to model reversible bimolecular reactions via diffusion encounter.
It should be mentioned a similar model was developed using a Master Equation
to derive modified reaction rates for simulations of the reaction diffusion master equation (RDME) \cite{fange2010stochastic}.
Nonetheless, the methodology follows a different logic; they assume the reaction rate 
in the Master Equation as unknown, and they use the continuous theory to derive an appropriate rate for some specific discretization.
In our work, we chose fixed reaction probability, and we showed the condition under which this recovers 
the continuous theory. We also provide the parameters of the continuous model in terms of the discrete probabilities,
which provides insight into the meaning of these quantities. In addition, the relation we obtained between 
the rate and the discretization parameters is simpler, and they don't extend their model to the cases discussed 
in the following Sections.

\subsection{Reversible reaction-diffusion with unbinding radius}\label{sec:ur}
In the previous model with reversible reaction, we assumed the $B$ particles are associated and dissociated in the same reaction shell
corresponding to $r=\sigma$. However, the probability of reacting/dissociating can be distributed along different spatial points and 
not only on a specific boundary. In the stochastic matrix (\ref{reactmat_rev}), the term $P_u$ can be placed in any of the columns 
in the first row. For instance, if this term is collocated in the $n^{th}$ column of the matrix, when the $B$ molecule is dissociated, 
it would be placed in the sphere $i=n$ with unbinding radius $\sigma_u=\sigma + n\delta r$ (see Figure \ref{fig:shells}). The binding 
process occurs in the boundary $i=0$ with $r=\sigma$, so we could say there is a binding radius $r=\sigma$ and an unbinding radius 
at $r=\sigma_u$. A new question now arises: if we consider an ensemble of these systems at thermodynamic equilibrium, is detailed 
balance satisfied?

\begin{figure}[t!]
    \centering
    \includegraphics[width=0.47\columnwidth]{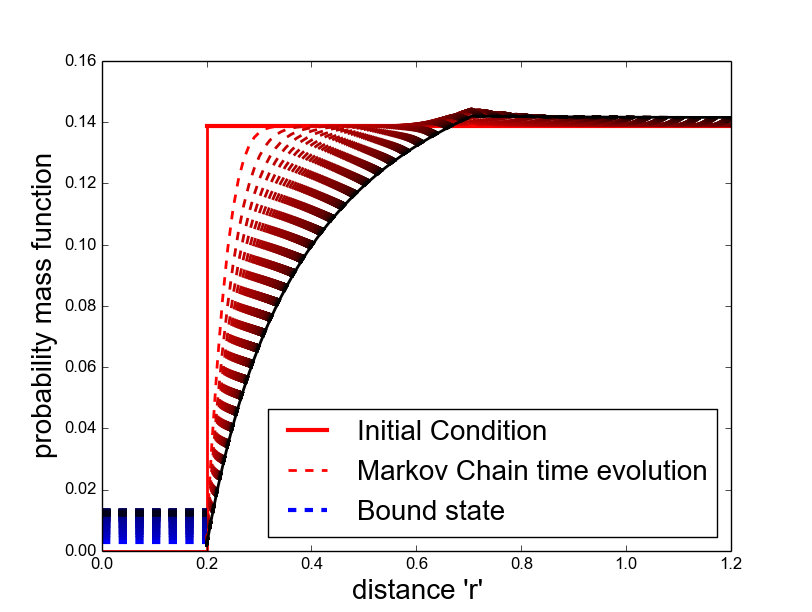} 
    \caption{Convergence of the discrete model to its steady state for the reversible case with an unbinding radius. The initial condition 
    is a uniform distribution in the unbound state and zero on the bound state, and it is plotted as a red line. 
    The dashed lines represent the convergence in time to the steady state of 
        the Markov chain from $t=0$ to $t=1$ taken every 100 time steps; the darker
        lines correspond to longer times. (a) Parameters used were: $\delta r = 0.01$, $\delta t = 0.0001$, $D = 0.1$, 
        $\tilde{\kappa}(\sigma) = 6000.0$, $\mu = 50$, $\sigma = 0.2$, $\sigma_u = 0.7$, 
        and 100 shells. The bound state $\pi_b$ value at different time steps is shown as an horizontal dashed line from $r=[0,\sigma]$. The final
        steady state is emphasized as a black continuous curve; as the slope from the right of $r=\sigma_u$ is zero, we know the net flux 
        for $r>\sigma_u$ is zero, as expected. Detailed balance is not satisfied between $r=\sigma$ and $r=\sigma_u$}
        \label{fig:UR_conv} 
\end{figure}

Detailed balance tells us that in a chemical kinetic system at equilibrium every elementary reaction is balanced by its 
reverse reaction $\pi_{i} P_{i \rightarrow j}=\pi_{j} P_{j \rightarrow i}$ \cite{lewis1925new}, where $P_{a \rightarrow b}$ is 
the transition probability from state $a$ to $b$. At equilibrium, detailed balanced will not be satisfied inside the 
shells $i=0$ and $i=n$ since we have an open system with a source and a sink. In these shells, we actually observe a NESS, like 
the one from Section \ref{sec:irrRD}. Nonetheless, we scaled the unbinding rate so the net flux coming out of shell $i=n$ is 
zero, so detailed balance will be satisfied everywhere else between shell $i=n$ and the outermost shell. This might seem at 
first confusing since detailed balance is based on chemical kinetics; however, the concept can be easily extended to diffusion 
and random walks. As we know that as long as there are no flux conditions in the boundaries of a system in steady state, detailed 
balance must be satisfied everywhere. As there is no flux condition at shell $i=n$ and at infinity (or at the outermost shell), 
detailed balance is satisfied between those two boundaries. This is indeed the reason why the particle-based stochastic 
reaction-diffusion algorithms that introduce an unbinding radius work. Introducing the unbinding radius sacrifices 
accuracy in the local region around $r=\sigma$, but it allows fast and accurate 
simulations \cite{andrews2004stochastic,del2014fluorescence} on slightly larger scales. In Figure \ref{fig:UR_conv}, we 
can observe the time convergence of the probability mass function to the steady stated for this case. In contrast with 
Figure \ref{fig:rev_conv}, we can see between $r=\sigma$ and $r=\sigma_u$ the solution has non-zero flux; this is the 
region where detailed balance will not be satisfied. However, the solution will be accurate everywhere else, 
where detailed balance is satisfied.

\subsection{Radial random walk with spherical symmetry under a potential}\label{sec:diffpot}
Assuming the random walk is influenced by a smooth interaction potential $U(r)$, we can modify the original transition probabilities in 
Eq. \ref{probdiff} by including the interaction potential term as
\begin{align}
\begin{split}
p_i=\delta t\left(\fr{D}{\delta r^2} - \fr{D}{r_{i-1}\delta r} 
                + \fr{\beta D}{4\delta r^2}\left[U_{i+1} - U_{i-1}\right]\right), \\
q_i=\delta t\left(\fr{D}{\delta r^2} + \fr{D}{r_{i+1}\delta r}
        - \fr{\beta D}{4\delta r^2}\left[U_{i+1} - U_{i-1}\right]\right),
\end{split}
\label{probdiff2}
\end{align}
where $\beta=1/k_B T$, with $k_B$ the Boltzmann constant and $T$ the temperature. 
With these new values of $p_i$ and $q_i$ in the matrix \ref{reactmat}, we can rewrite the $i^{\mathrm{th}}$ equation of (\ref{KFE}) as,
\begin{align}
 \pi_i&^{t+1} = \pi_i^{t} + \delta t D \left[ \fr{\pi_{i-1}^{t} -2\pi_i^{t} +\pi_{i+1}^{t}}{\delta r^2} \right] - 
        \delta t \fr{2D}{r_i}\left[\fr{\pi_{i+1}^{t} - \pi_{i-1}^{t}}{2\delta r}\right] \nonumber \\
               & + \delta t \fr{D}{\delta r}\left[\fr{\pi_i^{t}}{r_i - \delta r} - \fr{\pi_i^{t}}{r_i + \delta r}\right] 
  +\delta t \fr{\beta D}{2 \delta r} \left[\pi_{i+1}^t\left( \fr{U_{i+2} - U_i}{2\delta r}\right) - 
  \pi_{i-1}^t\left( \fr{U_{i} - U_{i-2}}{2\delta r}\right)\right].
  \label{eq:discpot}
\end{align}
Dividing by $\delta r$ and taking the limit as before we obtain
\begin{gather}
 \fr{\p \Pi(r,t)}{\p t} = D\fr{\p^2 \Pi(r,t)}{\p r^2} - \fr{\p}{\p r}\left( \fr{2D}{r} \Pi(r,t)\right) + 
 \beta D\fr{\p}{\p r}\left[\Pi(r,t)\fr{\p U(r)}{\p r}\right],
 \label{eq:FPEdr2}
\end{gather}
with second order accuracy in space.
Employing once again the fact that $\Pi(r,t) = 4\pi r^2 f(r,t)$, we recover the Smoluchowski equation under a potential \cite{szabo1989theory},
\begin{align}
 \fr{\p f(r,t)}{\p t} &= \fr{D}{r^2} \fr{\p}{\p r} \left(r^2 \left[ \fr{\p f(r,t)}{\p r} + \beta f(r,t)\fr{\p U(r)}{\p r}\right]\right) \\
 &= \fr{D}{r^2} \fr{\p}{\p r} \left(r^2 e^{-\beta U(r)} \fr{\p}{\p r} e^{\beta U(r)}f(r,t) \right).
 \label{eq:heat2}
\end{align}
We have not talked about the boundary yet. This will be a very subtle issue as we will comment on further. For the time being,
lets assume the boundary is given as that of the matrix (\ref{reactmat}). In such case the first difference equation of the Kolmogorov
forward equation yields,
\begin{align*}
 \pi_0^{t+1} = &\pi_0^{t} + \fr{\delta t D}{\delta r ^2} \left[ \pi_{1}^{t} -2\pi_0^{t} + 
                \pi_0^{t}\left(1-\fr{\delta r ^2}{D}\fr{\kappa}{4\pi r_0^2 \delta r} \right) \right]  
      -\delta t \left[\fr{D}{r_0}\fr{\pi_{1}^{t}}{\delta r} + \fr{D}{\delta r} \left(\fr{\pi_0^{t}}{r_0 + \delta r}\right)\right] +\\
      &\fr{\delta t \beta D}{2 \delta r} \left[\pi_1^t\left(\fr{U_2 - U_0}{2\delta r}\right) +
                                                \pi_0^t\left(\fr{U_1 - U_{-1}}{2\delta r}\right)\right].  \\[2mm]
\end{align*}
Once again, in order to find the boudary condition, we rewrite this equation in the same form than Eq. \ref{eq:discpot}
by introducing the ghost cell $\pi_{-1}^{t+1}$. This will require introducing values for the potential at two ghost cells $U_{-1}$ and $U_{-2}$,
\begin{align*}
   \pi_0^{t+1} = &\pi_0^{t} + \delta t D \left[ \fr{\pi_{1}^{t} -2\pi_0^{t} +\pi_{-1}^{t}}{\delta r^2} \right] - 
                \delta t \fr{2D}{r_0}\left[\fr{\pi_{1}^{t} - \pi_{-1}^{t}}{2\delta r}\right] + 
                 \delta t \fr{D}{\delta r}\left[\fr{\pi_{0}^{t}}{r_0 - \delta r} - \fr{\pi_0^{t}}{r_0 + \delta r}\right] +\\
                &\delta t \fr{\beta D}{2 \delta r} \left[\pi_{1}^t\left( \fr{U_{2} - U_0}{2\delta r}\right) - 
                \pi_{-1}^t\left( \fr{U_{0} - U_{-2}}{2\delta r}\right)\right],
\end{align*}
where $U_{-1}$ and $U_{-2}$ are usually well defined since the boundary where $i=0$ corresponds to $\sigma>0$ and $\delta r$ is small. 
In the case where these are undefined, we can always use articial values to match the right derivative and curvature 
of the potential at the boundary. This equation divided by $\delta r$ in the limit $\delta_t \rightarrow 0$ and $\delta_r \rightarrow 0$ is 
again reduced to the appropriate Fokker-Planck equation, given by Eq. \ref{eq:FPEdr2}. However, analogously as before,
$\pi_{-1}^t$ needs to satisfy,
\begin{gather*}
\pi_0^t - \fr{\delta r \kappa}{4\pi Dr_0^2}\pi_0^t +\fr{\beta}{4}\pi_0^t(U_1 - U_{-1}) 
= \pi_{-1}^t + \delta r\fr{\pi_0^t }{r_{-1}} + \delta r\fr{\pi_{-1}^t }{r_{0}} - \fr{\beta}{4} \pi_{-1}^t(U_0 - U_{-2}),
\end{gather*}
which in the continuous limit becomes
\begin{gather*}
 \left. \fr{\p \Pi(r,t)}{\p r}\right|_{r=\sigma} = \fr{\kappa}{4\pi\sigma^2D}\Pi(\sigma,t) + \fr{2\Pi(\sigma,t)}{\sigma} 
 -\beta \left. \fr{\p U(r)}{\p r}\right|_{r=\sigma} \Pi(r,t),
\end{gather*}
with first order accuracy in space. Using again Eq. \ref{eq:BMtrans}, we can rewrite it as
\begin{gather*}
 4 \pi \sigma^2 D \left. \fr{\p f(r,t)}{\p r} \right|_{r=\sigma} = \kappa f(\sigma,t) 
 - 4 \pi \sigma^2 D \beta \left. \fr{\p U(r)}{\p r}\right|_{r=\sigma}f(\sigma,t) \\[2mm] 
 \Rightarrow \ \ \ 
  4 \pi \sigma^2 D e^{-\beta U(\sigma)} \left[ \fr{\p e^{\beta U(r)}f(r,t)}{\p r} \right]_{r=\sigma} = \kappa f(\sigma,t),
\end{gather*}
which is the Collins-Kimball-Debye boundary condition \cite{shoup1982role}. If we would have used matrix (\ref{reactmat_rev}) instead, 
we would have obtained the back-reaction boundary condition in Eq. \ref{eq:backreac_BC} with the addition of the potential factors. 
However, there is the subtlety of how to define the bound state that we discussed in Section \ref{sec:boundun}. Does it depend on the 
distance between $A$ and $B$ molecules, or does it depend on the state? According to our previous discussion, it will depend on the 
physical problem at task. 

\begin{figure}[t!]
    \centering
%    \begin{subfigure}[t]{0.4\textwidth}
%         \centering
        \includegraphics[width=0.47\columnwidth]{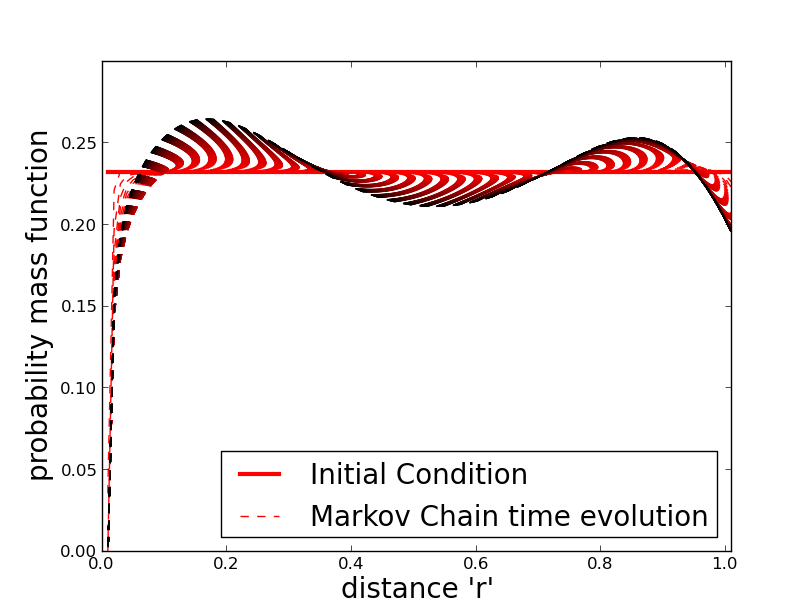} 
%         \caption{}
%    \end{subfigure}
%   \begin{subfigure}[t]{0.4\textwidth}
%         \centering
        \includegraphics[width=0.47\columnwidth]{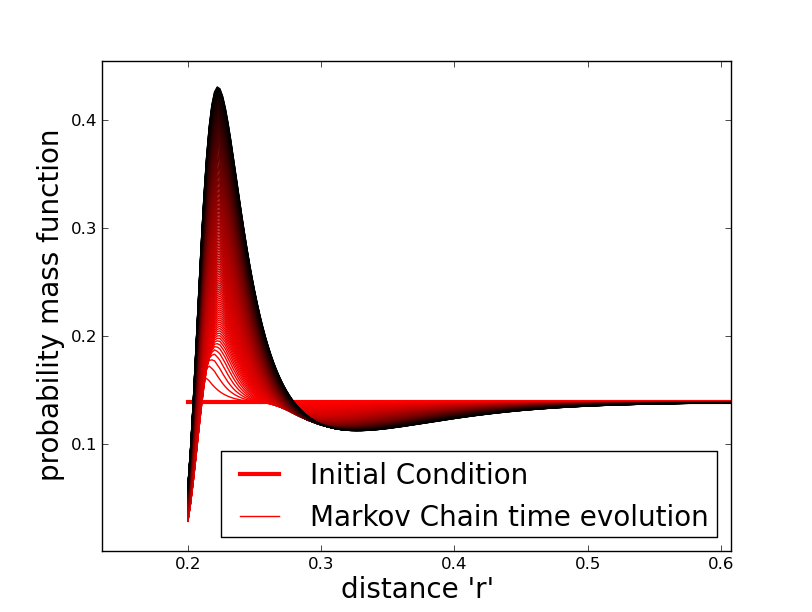} 
%         \caption{}
%    \end{subfigure}
    \caption{Convergence to steady state for the diffusion under a force field case for two potentials. The initial condition
        for the discrete Markov model, which is a uniform distribution, is plotted as a thick red line. 
        The thin lines represent the convergence in time to the steady state of 
        the Markov chain from $t=0$ to $t=0.2$ taken every 10 time steps; the darker
        lines correspond to longer times. (a) Plot using potential $U_{(r)} = U_{K}(r)$.
        The parameters used were $V_0 = 0.0001$, $r_m=0.5$, $A=2.0$, $\delta r = 0.01$, $\delta t = 0.0001$, $D = 0.1$, 
        100 shells and $r_0=0$.(b) Plot using a Lennard-Jonnes type potential $U(r) =U_{LJ}(r)$.
        The parameters were $V_0 = 0.00005$, $r_m=0.22$ $\delta r = 0.002$, $\delta t = 2E-5$, $D = 0.1$, 
        500 shells and $r_0=0.2$.}
        \label{fig:kram_conv}
\end{figure}

In Figure \ref{fig:kram_conv}, we plot the convergence to the steady state for the two potential functions,
\begin{align*}
 U_K(r) &= V_0\left[(A(r-r_m))^4-(A(r-r_m))^2\right], \hspace{5mm} \mathrm{and} \\
 U_{LJ}(r) &= V_0\left[\left(\fr{r_m}{r}\right)^{12}-2\left(\fr{r_m}{r}\right)^6\right].
\end{align*}
These correspond to the potential in Kramers original work and a Lennard-Jonnes potential. We observe
the steady state distribution exhibits bistability, even when using the Lennard-Jonnes potential due to the
geometrical drift. In this case, the reversible reaction can be modeled without the need of a boundary, and the state
of $B$ can be stated as bound or unbound depending on which side of the separatrix of the potential it is. Note
this still matches both the distance and state definition mentioned in Section \ref{sec:boundun}. However, if the 
potential doesn't only depend on $r$ but on more variables, the distance and state definition might not provide the same answer. 

Once again, as we have a steady state and zero flux boundaries, detailed balance must be satisfied,
and the equilibrium state satisfies the Gibbs-Boltzmann distribution. We should also mention an alternative 
discretization for discontinuous potentials was introduced in \cite{wang2003robust}.
Their discretization is only performed in one dimension and does not implement reversible reacting boundaries. 
In the future, it would be interesting to explore how to extended their to include reversible reaction boundaries.
Another very relevant work was done in \cite{krissinel1996spherical,pines1988geminate}, where a similar discretization as a Markov
chain is given; however, the emphasis is on providing a numerical solution of the Smoluchowski equation instead of a stochastic 
framework that allows particle based simulations in conjunction with probability mass function dynamics. 
 
\subsection{Extension to multiple molecules} 
\label{sec:ext}
The aim of the present work is to unify theories and algorithms
and to provide a deeper understanding of stochastic reaction-diffusion processes, by using 
a simple intuitive model. Nonetheless, we will offer some suggestions on its value in computations,
in particular, its extension to multiple-molecule simulations. One of the key advantages of this 
formulation, in comparison to other algorithms, is that it can 
handle multiple $B$'s in particle-based simulations naturally on the same grid, and it 
doesn't need to decouple a system into multiple two-body problems. Of course, once a 
$B$ has been absorbed a new $B$ cannot be absorbed until the previous one is 
dissociated. In this sense, our model already handles multiple molecules.

The real challenge is when one also has multiple $A$'s. If we assume $A$'s are diluted macromolecules, 
there are several possibilities to extend this model that would be worthwhile to explore. One possibility 
is to create non-overlapping truncated spherical grids around each $A$ molecule. Each one of this molecules 
diffuses along with its grid following a random walk with diffusion coefficient $D_A$. In the same 
manner, $B$ molecules diffuse on open space following a random walk with coefficient $D_B$, unless 
they diffuse into one of the outermost spherical shells or if they are already inside one. 
If a $B$ molecule diffuses into one of the outer shells, it will undergo diffusion inside the spherical shells 
following the model from the previous sections. If the molecule 
escapes the outermost shell of the grid, it will return to diffuse freely with the direction of exit 
chosen randomly from a uniform distribution. Complications will arise if at some time step two 
grids of $A$ overlap. In this case, we would need to calculate new non-overlapping grids for the 
two $A$'s, which will force a smaller time step. However, as $A$ is dilute this should not happen 
very often. In the case where $A$ molecules are binding sites fixed in space, this could be an 
efficient and accurate approach.

Another possible implementation is to precompute the probability distribution function 
in a large and high-resolution grid and save these into a lookup table. The table 
could then be used as the reference solution to compute the 
probabilities in simulation algorithms like eGFRD or FPKMC. The difference is that this solution
uses the back-reaction boundary condition and not the partially absorbing one. This could 
speed up existing algorithms since the partially absorbing boundary requires modeling every dissociation
event while the back-reaction boundary does not. This model provides an easy and intuitive 
alternative to compute lookup tables for third-parties interested in developing their own simulations,
and it also encourages reproducibility. It would be interesting to compare the lookup tables obtained
in our discrete model with those of other simulation packages. Implementation of these 
extensions and the comparison studies between different algorithms is left as future work.

\section{Discussion} \label{sec:disc}
We developed a discrete time/space stochastic model for 
bimolecular chemical reaction via diffusion encounter.  The
model converges to all the well known classical results 
of continuous irreversible diffusion-influenced 
reaction theory. It also allows diffusion under a force field, like in 
the theories of Debye and Kramers.

The significance of this formulation is for the case of reversible reactions, 
where there have been extensive discussions on the best approach to model the process. One of the main models
is given by the usual diffusion equation, Eq. \ref{eq:smol2}, coupled with the back-reaction 
boundary condition from Eq. \ref{eq:backreac_BC}. The main issue with this model, however, is that
the back-reaction boundary condition is so complicated that it obscures the underlying stochastic 
process and consequently an accurate and transparent particle-based simulation. The discrete model 
presented in the present paper was inherently constructed as a stochastic process using simple notions 
of discrete association and dissociation reactions. In the continuous limit, it recovers the reversible
diffusion-influenced theory, providing a clear description of 
the underlying stochastic process, which is hard to grasp in the continuous limit due to the nature of 
PRBM as shown in the Appendix \ref{sec:appB}. In addition, it
provides the continuous model parameters as a function of the jump and reaction probabilites, which 
elucidates the probabilistic interpretation of the parameters in the classical theory. 
It also allows straightforward simple Markovian algorithms to compute particle-based simulations and  
to compute the probability distribution, ensuring consistency between the two as well as conservation 
of probability. 

In the continuous limit, we recover a more extensive version
of the original back-reaction model, as in the Eqs. \ref{eq:BR_newmodel1}, \ref{eq:BR_newmodel2} and 
\ref{eq:BR_newmodel3}. This result unifies the back-reaction boundary approach for 
reversible diffusion-influenced reactions with our discrete model. Furthermore, as we modeled the 
dissociation process following an exponential waiting time, the previous result also shows consistency 
between the reversible diffusion-influenced reaction theory and the approaches taken by other 
simulation algorithms, like eGFRD and FPKMC \cite{takahashi2010spatio,donev2010first}.

The model also allows an immediate implementation of an unbinding radius for the dissociation process, 
and it provides the correct scaling for the dissociation rate 
in terms of the diffusion controlled association rate. 
This approach can be used by other simulation algorithms to speed up simulations in a similar manner  
as done in Smoldyn \cite{andrews2004stochastic}. However, high accuracy and detailed balance are lost 
in the region inside the unbinding radius.

One disadvantage is the high accuracy of the discrete model 
causes numerical computations to be slow. Additionally, the model currently only
works for one $A$ and one or multiple $B$'s; however, we provided a guideline for possible implementations
with multiple $A$'s. These are left as future work. Other 
future directions are to implement extensions for volume reactivity models \cite{doi1976stochastic} and
more complicated scenarios where a reaction is not only weighted as a function of space but also 
other variables. Regardless of its computation capabilities, the model itself provides a lot 
of insight on the modeling of reversible reactions, and it unifies different approaches by 
providing an underlying common stochastic framework. The discrete stochastic nature of this 
model establishes a research guideline that could lead to more robust computational 
solutions of complex models, where else continuous models might become increasingly 
convoluted, obscure or even intractable. 

\section*{Acknowledgments} \label{sec:ack}
We thank Drs. Noam Agmon, Elliot Elson, Sam Isaacson and
Attila Szabo for helpful discussions and 
encouragements over the course of this work. M. J. R. 
acknowledges support from National Science 
and Technology Council of Mexico (CONACyT).  H. Q. acknowledges 
partial support from NIH grant R01GM109964 (PI: Sui Huang).

% BibTeX users please use

% \bibliographystyle{plain}
% \bibliographystyle{acm}
\bibliographystyle{myabbrv}
\bibliography{mauricio_ref}

\appendix
\numberwithin{equation}{section}

\section{Second order accuracy in no-flux boundaries} 
\label{app:secord}
In the discretization of Section \ref{sec:randwalk}, we mentioned that we have second order accuracy in all the inner points.
However, the boundary condition we obtain from Eq. \ref{eq:refbc} is only first order accurate. Lets check this, substracting 
Eq. \ref{eq:refbc} from $\pi_0^t$ and dividing by $\delta r$ we obtain,
\begin{align*}
\fr{\pi_0^t- \pi_{-1}^t}{\delta r} &=
\pi_0^t\fr{1}{\delta r}\left[1-\fr{\left(\fr{1}{\delta r}-\fr{1}{r_{-1}}\right)}{\left(\fr{1}{r_0} + \fr{1}{\delta r}\right)}\right]
=\fr{\pi_0^t}{r_0^2} \fr{2r_0 - \delta r}{1-\left(\delta r/r_0\right)^2} \\
&= \pi_0^t \fr{2r_0 - \delta r}{r_0^2}\sum_{i=0}^{\infty} \left(\fr{\delta r}{r_0}\right)^{2i}
=\fr{2 \pi_0^t}{r_0}\left[1 + O(\delta r)\right].
\end{align*}
Following standard finite difference theory \cite{randysbbook}, we take the limit as $\delta r \rightarrow 0$ and recover the zero 
flux boundary condition $\left. \fr{\p \Pi(r,t)}{\p r} \right|_{r=r_0}= \fr{2\Pi(r_0,t)}{r_0}$ to first order. In general, this would 
not be desirable when discretizing a partial differential equation for simulation since the first order error would 
be propagated to the rest of the solution. Nonetheless, it is not clear how relevant the accuracy at the boundaries is when interpreting 
the discretization as jump probabilities in a Markov chain. Once again, the actual stochastic process is very clear in the discrete 
scenario since we are modeling a discrete stochastic process instead of the discretization of a continuous stochastic process. We 
also know that the reactions and the no-flux boundaries are modeled appropriately in the
discrete process. How relevant is that we recover the 
continuous version of the boundary conditions with high accuracy is up for debate. The important part is that we do recover them 
showing consistency between the discrete and continuous models. In theory, we could try
to modify the rates to obtain a second order discretization. However, it seems it would not be possible to write this 
discretization as a Markov chain anymore.

In the case of no-flux boundary conditions, there is an alternative approach that will yield second order accuracy. 
Lets concentrate on the steady state $\mathbf{\pi}^{ss}$, which we know satisfies detailed balance in the 
boundary $\pi_0^{ss} q_0 = \pi_1^{ss} p_1$. The detailed balance condition for the problem in Section \ref{sec:randwalk} 
can be written as,
\begin{align}
\pi_{0}^{ss}\left(\fr{1}{r_0} + \fr{1}{\delta r}\right) = \pi_1^{ss} \left(\fr{1}{\delta r}-\fr{1}{r_{-1}}\right).
\label{app:detbal}
\end{align}
Substracting Eq. \ref{eq:refbc} from $\pi_1^{ss}$ and dividing the result by $2 \delta r$, we obtain
\begin{align*}
 \fr{\pi_1^{ss}- \pi_{-1}^{ss}}{2 \delta r} =\fr{1}{2 \delta r} \left[\pi_1^{ss} - 
 \pi_0^{ss}\fr{\left(\fr{1}{\delta r}-\fr{1}{r_{-1}}\right)}{\left(\fr{1}{r_0} + \fr{1}{\delta r}\right)}\right].
\end{align*}
Using the detailed balance condition of Eq. \ref{app:detbal}, we can write the right hand side in terms of $\pi_0^{ss}$, which yields
\begin{align*}
 \fr{\pi_1^{ss}- \pi_{-1}^{ss}}{2 \delta r} &=\fr{\pi_0^{ss}}{2 \delta r} \left[\fr{\left(\fr{1}{\delta r}+
 \fr{1}{r_{1}}\right)}{\left(\fr{1}{\delta r} - \fr{1}{r_0}\right)} - \fr{\left(\fr{1}{\delta r}-\fr{1}{r_{-1}}\right)}{\left(\fr{1}{r_0} + 
 \fr{1}{\delta r}\right)}\right] \\
 &=\fr{2 r_0\pi_0^{ss}}{r_0^2-\delta r^2} = \fr{2\pi_0^{ss}}{r_0} \sum_{i=0}^{\infty} \left(\fr{\delta r}{r_0}\right)^{2i} 
 = \fr{2\pi_0^{ss}}{r_0} \left[1 + O(\delta r^2)\right].
 \end{align*}
In the limit $\delta r \rightarrow 0$, it satisfies the no flux boundary condition with second order accuracy. This result 
is a consequence of the fact that our discretization can be interpreted as a Markov chain with detailed balance. 
The downsides are that this level of accuracy is only attained in the steady state solution and that it is not 
applicable for reactive boundaries.
 
\section{Partially reflected Brownian Motion (PRBM) as a limit of a random walk} \label{sec:appB}
Consider a random walk moving in one dimensional space. This can be described as a discrete time Markov chain, where 
the state $i$ refers to its position $x_i=i\delta x$. Let the probability of jumping from state $i$ to $i+1$ 
be $q_i$ and from state $i$ to $i-1$ be $p_i$, such that $p_i+q_i \le 1$ and $1-(p_i+q_i)$ is the probability of 
not jumping. For a purely reflective boundary 
at $i=0$, the only possibility is that the probability from jumping from state 
$i=0$ to the absorbed state is $p_0 = 0$. Analogously, for the purely absorbing boundary 
at $i=0$, it is required that $p_0 = 1$. In the same manner, we can easily
think of the partially absorbing boundary condition at $i=0$, where the probability of being absorbed 
is $p_0 =\epsilon$, with $\epsilon \in (0,1)$. 

In order to study this random walk, lets assume $\delta x$ is fixed, so in order 
to satisfy that $p_i+q_i \le 1$, we need $\delta t \le \delta x^2/(2D)$. 
We would like to relate this random walk to Brownian Motion, 
so we need to assign correct values to $q_i$ and $p_i$. This is analogous to what we did in Section \ref{sec:discretemodel}. 
The jump probabilities elsewhere outside the boundary should be $q_i=p_i=D\delta t/\delta x^2$. Following the same structure of
the Markov chain from Eq. \ref{KFE} with Eq. \ref{reactmat}, the $i^{th}$ equation yields 
\begin{align}
 \fr{\pi_i^{n+1} - \pi_i^n}{\delta t} = D\left[\fr{\pi_{i+1}^n - 2\pi_{i}^n + \pi_{i-1}^n}{\delta x^2}\right],
 \label{eq:fdPRBM}
\end{align}
which in the continuous limit recovers the one dimensional Brownian Motion $\p \Pi/ \p x = D \p^2 \Pi/\p x^2$.
We also would like to recover the partially absorbing boundary condition, so we write the first equation in the Markov chain,
\begin{align*}
 \fr{\pi_0^{n+1} - \pi_0^n}{\delta t} = \fr{D}{\delta x^2}\pi_1^n-\fr{D}{\delta x^2}\pi_0^n - \frac{\epsilon}{\delta t} \pi_0^n,
\end{align*}
where we used that the probability of staying at shell $i=0$ is $1-(q_0+p_0)=1-(D\delta t/\delta x^2 + \epsilon)$.
In order for this equation to have the same form as Eq. \ref{eq:fdPRBM}, we need to introduce the ghost 
shell $\pi_{-1}$ (as explained in Section \ref{sec:randwalk}), which satisfies
\begin{align*}
 \epsilon \pi_0^n = D\delta t \fr{\pi_0^n - \pi_{-1}^n}{\delta x ^2}.
\end{align*}
In order to be able to recover the partially absorbing boundary 
condition $D\p \Pi/\p x |_{x=x_0} = \kappa \Pi(x_0,t)$, we need to set $\epsilon = \kappa \delta t/\delta x$. 
It is now straightforward to write $\kappa$ and the Diffusion coefficient in terms of the jump 
probabilities in the discrete model by summing $p_i$ and $q_i$,
\begin{align}
D = \fr{\delta x^2}{2\delta t}, \hspace{5mm} \kappa =  \fr{\delta x}{\delta t}\epsilon.
\label{eq:Dkdisc}
\end{align}
Note $\epsilon = 0$ corresponds to a purely reflective boundary, which means $\kappa = 0$, so 
$\p \Pi/\p x|_{x=x_0} = 0$. The purely absorbing boundary corresponds to $\epsilon \rightarrow 1$. 
As shown in Section \ref{sec:randwalk}, $\delta t \le \delta x^2/(2D)$, so $\kappa=\delta x\epsilon/ \delta t \ge 2D \epsilon/\delta x$.
Then, as $\delta x \rightarrow 0 \Rightarrow \kappa \rightarrow \infty$, and the boundary 
condition becomes $\Pi(x_0,t) = 0$. As expected, this matches the limiting behavior 
from Collins and Kimball's original work \cite{collins1949diffusion}. 
 
A very subtle issue is in the continuous limit of the probability of partial 
absorption $\epsilon =  \kappa \delta t /\delta x$ with $\kappa$ constant. 
Again as $\delta t \le \delta x^2/(2D)$, then  $\epsilon \le \kappa\delta x/ (2D)$.
This means that in the continuous limit $\delta x \rightarrow 0$, the probability of 
partial absorption goes to zero, $\epsilon\rightarrow 0$. However, $\epsilon = 0$ 
means a purely reflective boundary, not a partially absorbing one, so the continuous limit of partially absorbing boundaries 
can be confusing, especially when trying to interpret them in particle-based simulations. 
The works \cite{grebenkov2006,burdzy-chen-2008} elaborate on this issue and state that partially reflected Brownian Motion (PRBM)
should be understood as purely reflective Brownian Motion conditioned to stop at a random moment given in terms of the local time 
process. In more intuitive terms, given a counting process $\mathcal{L}_t$ that counts the number of hits of a particle to the 
reflective boundary, the particle is killed (absorbed) when $\mathcal{L}_t\ge \chi$, where $\chi$ is an independent 
exponentially distributed random variable. It is in this sense that
PRBM is best understood as the limit of a random walk. Furthermore, the parameters $D$ and 
$\kappa$ have a clear interpretation in terms of the jump and partial absorption probabilities. This result provides 
a more insightful understanding of the parameters even in the continuous model.

Also note the probability of absorption is a function
of $\delta t$, $\epsilon = \kappa \delta t /\delta x$. Assuming we chose the largest possible value of 
$\delta t = \delta x^2/(2D)$, we have $\epsilon = \kappa \delta x /(2D)$. However, the fact that 
$ 0\le \epsilon \le 1$ limits our choice of $\delta x$. This bring into light that this choice of $\epsilon$
might not be the most appropriate one for a Markovian model. On the other hand, the 
probability $\epsilon = \kappa \delta t /\delta x$ could be understood as a first order 
approximation when sampling from an exponential waiting time,
\begin{align}
 1 - e^{-\kappa \fr{\delta t}{\delta x}} \approx \kappa \fr{\delta t}{\delta x}.
 \label{eq:exprte}
\end{align}
This is basically emphasizing the fact that the reaction process at the boundary is Poissonian.
Using $\epsilon = 1 - e^{-\kappa \delta t/\delta x}$ allows a more accurate choice for the 
discrete partial reaction probability since it always satisfy $ 0\le \epsilon \le 1$. In this case,
the discrete parameter $\kappa$ is given by $\kappa =  -\fr{\delta x}{\delta t}\log{[1-\epsilon]}$.
It also gives the upper bound for the $\kappa$ parameter in a given discretization of the diffusion process.
As $p_0+q_0 \le 1$, then $\epsilon + D\delta t /\delta x^2 \le 1$, so 
$\kappa \le -\frac{\delta x}{\delta t}\log\left[D\frac{\delta t}{\delta x^2}\right]$; the equality corresponds to
the purely absorbing case. 
Note that using the result from Eq. \ref{eq:exprte} in the Markov model does not affect any of the previous limiting results.
As a matter of fact, all the models in Section \ref{sec:discretemodel} can use this result without affecting the
continuous limit behavior. It should also be noted this result would have never been obtained by looking for an 
accurate discretization of the continuous models; it is only obtained by looking for consistency with the
underlying stochastic process.

\end{document}